\documentclass[twocolumn,apj,twocolappendix,numberedappendix]{openjournal}

\usepackage{xcolor}
\usepackage[utf8]{inputenc}
\usepackage[english]{babel}

\usepackage{color}
\definecolor{linkcolor}{rgb}{0.0,0.3,0.5}

\usepackage{hyperref}
\hypersetup{
    unicode, 
    colorlinks=true,
    linkcolor=linkcolor,
    citecolor=linkcolor,
    filecolor=linkcolor,
    urlcolor=linkcolor,
}

\DeclareGraphicsExtensions{.bmp,.png,.jpg,.pdf}
\usepackage{orcidlink}

\usepackage{graphicx}
\usepackage{amsmath}
\usepackage{listings}
\usepackage{xspace}

\newcommand{\synthesizer}{\textsc{Synthesizer}\xspace}
\newcommand{\synth}{\textsc{Synthesizer}\xspace}
\newcommand{\unyt}{\textsc{unyt}\xspace}
\newcommand{\emodel}{\texttt{EmissionModel}\xspace}
\newcommand{\emodels}{\texttt{EmissionModels}\xspace}
\newcommand{\grid}{\texttt{Grid}\xspace}

\newcommand{\app}[1]{Appendix~\ref{sec:#1}}

\newcommand{\fig}[1]{Figure~\ref{fig:#1}}
\renewcommand{\sec}[1]{Section~\ref{sec:#1}}
\newcommand{\tab}[1]{Table~\ref{tab:#1}}

\newcommand\blfootnote[1]{%
  \begingroup
  \renewcommand\thefootnote{}\footnote{#1}%
  \addtocounter{footnote}{-1}%
  \endgroup
}

\graphicspath{ {./images/} }

\begin{document}

\title{Synthesizer: a Software Package for Synthetic Astronomical Observables \vspace{-3em}}

\author{Christopher C. Lovell$^{1,2,3 \star \dagger}$ \orcidlink{0000-0001-7964-5933}}
\author{William J. Roper$^{4 \star \dagger}$\orcidlink{0000-0002-3257-8806}}
\author{Aswin P. Vijayan$^{4 \star \dagger}$\orcidlink{0000-0002-1905-4194}}
\author{Stephen M. Wilkins$^{4 \star \dagger}$\orcidlink{0000-0003-3903-6935}}
\author{Sophie Newman$^{3}$\orcidlink{0009-0001-3422-3048}}
\author{Louise Seeyave$^{4}$\orcidlink{0000-0000-0000-0000}}

\blfootnote{$^{\star}$ Denotes principal authors with equal contribution.}
\blfootnote{$^{\dagger}$ Email: \href{mailto:chris.lovell.astro@gmail.com}{chris.lovell.astro@gmail.com}, \href{w.roper@sussex.ac.uk }{w.roper@sussex.ac.uk }, \href{a.payyoor-vijayan@sussex.ac.uk}{a.payyoor-vijayan@sussex.ac.uk}, \href{s.wilkins@sussex.ac.uk}{s.wilkins@sussex.ac.uk}}

\affiliation{$^1$Kavli Institute for Cosmology, Madingley Road, Cambridge CB3 0HA, UK}
\affiliation{$^2$Institute of Astronomy, Madingley Road, Cambridge CB3 0HA, GB}
\affiliation{$^3$Institute of Cosmology and Gravitation, University of Portsmouth, Burnaby Road, Portsmouth PO1 3FX, UK}
\affiliation{$^4$Astronomy Centre, University of Sussex, Falmer, Brighton BN1 9QH, UK}

\begin{abstract}
We present \textsc{Synthesizer}, a fast, flexible, modular and extensible platform for modelling synthetic astrophysical observables.
\textsc{Synthesizer} can be used for a number of applications, but is predominantly designed for generating mock observables from analytical and numerical galaxy formation simulations.
These use cases include (but are not limited to) analytical modelling of the star formation and metal enrichment histories of galaxies, the creation of mock images and integral field unit observations from particle based simulations, detailed photoionisation modelling of the central regions of active galactic nuclei, and spectro-photometric fitting.
We provide a number of stellar population synthesis models, photoionisation code configurations, dust models, and imaging configurations that can be used `out-of-the-box' interactively.
The code can be used to quantitatively test the dependence of forward modelled observables on various model and parameter choices, and rapidly explore large parameter ranges for calibration and inference tasks.
We invite and encourage the community to use, test and develop the code, and hope that the foundation developed will provide a flexible framework for a number of tasks in forward and inverse modelling of astrophysical observables.
The code is publicly available at \href{https://synthesizer-project.github.io}{synthesizer-project.github.io}.
\end{abstract}

\begin{keywords}
    {
methods: numerical, galaxies: general, virtual observatory tools}
\end{keywords}

\maketitle

\section{Introduction}

There are many ways of comparing theoretical models of galaxy formation and evolution with observations.
One of the most common approaches is to fit the observations to derive constraints on resolved and integrated physical properties and parameters \citep{walcher_fitting_2011,conroy_modeling_2013}.
This procedure, known as spectral energy distribution (SED) fitting, or `inverse modelling', involves many simplifying assumptions \citep{pacifici_art_2023}, such as parameterised star formation histories, and screen models for dust attenuation, which can lead to biases in derived parameters, such as the galaxy stellar mass \citep[e.g.][]{lee_biases_2009,mitchell_how_2013,hayward_should_2015,lower_how_2020}.
Despite these uncertainties, the derived properties are often used to test phenomenological models, or compared directly with the predictions of numerical simulations in this \textit{physical} properties space \citep{somerville_physical_2015}.

Of increasing importance is the opposite approach, where `forward models' are used to predict the multi-wavelength electromagnetic emission from galaxies in numerical simulations \citep[e.g.][]{trayford_colours_2015,trayford_optical_2017,dave_mufasa:_2017,vijayan_first_2021}.
In this approach a unique mapping is provided from the physical to the observable space, rather than a non-unique inversion with some uncertainties \citep{torrey_synthetic_2015}, and allows for the inclusion of instrumental and observational effects, such as instrumental noise, point spread function, pixel effects and survey geometry.
A range of forward modelling techniques have been developed;
the predictions from such models can then be compared to observations directly in the \textit{observable properties} space, avoiding many of the simplifying assumptions and uncertainties in SED fitting, and providing an arguably clearer `apples-to-apples' comparison \citep{conroy_propagation_2010}.
However, it is important to emphasise that both forward-- and inverse--modelling approaches are important for understanding the uncertainties and biases in our models, and for deriving the maximum information and physical insight from observational data and theoretical models.

Many different components make up a galaxy and contribute to its emission.
Stellar populations produce multi-wavelength emission that is strongly dependent on their age and metallicity.
Young star forming regions photoionise their nebular birth clouds, leading to line and continuum emission. The interstellar radiation field can also ionise diffuse gas, while the cooling of this gas, driven by various elements and molecules, also produces line emission in the far-infrared (FIR).
Dust makes up a small fraction of the total baryon mass budget, but has an outsize effect on a galaxy's emission, leading to significant attenuation in the ultraviolet, and re-emission in the infrared and mm-regime.
Polycyclic Aromatic Hydrocarbons (PAHs) are organic molecules that predominantly contribute to the mid-infrared emission \citep{tielens_interstellar_2008}.
Actively accreting supermassive black holes at the centers of galaxies, known as Active Galactic Nuclei (AGN), can contribute to the emission across the whole electromagnetic spectrum, from the X-ray to the radio.
Enriched gas in the interstellar and intergalactic medium can create multiple absorption features in the composite continuum, and neutral gas can lead to UV absorption features.
In addition, multiple backgrounds, including the cosmic microwave and infrared backgrounds, can alter the physical properties of sources and attenuators, as well as contributing to the overall emission.

All of these sources and processes are required to accurately model the multiwavelength emission from a sufficiently high fidelity simulated galaxy.
A number of codes have been developed to model the emission from these various components, of varying sophistication and resulting computational expense.
The choice of method depends primarily on the content (dark matter only / hydrodynamic) and fidelity (mass and spatial resolution) of the numerical simulation they are being applied to, as well as whether the integrated or spatially resolved emission is required.

The emission from co-eval stellar populations is typically modelled by assuming an initial mass function (IMF) and a stellar population synthesis (SPS) model, that can include a range of different modelling assumptions, such as stellar multiplicity \citep{eldridge_binary_2017,stanway_re-evaluating_2018,byrne_dependence_2022}, or the contribution of rare stellar phases \citep[e.g.][]{maraston_evolutionary_2005}.
The aggregate star formation and metal enrichment history (SFZH) of the galaxy is then a composite of these co-eval populations.
This can be modelled using analytic prescriptions, which provide the integrated SFZH in a parametric (or functional) form \citep{iyer_reconstruction_2017,carnall_how_2019}, or through more flexible non-parametric (or non-functional) approaches \citep{leja_how_2019}.
Semi-analytic models, based on dark matter only numerical simulations, can model more complex star formation histories \citep{,baugh_primer_2006}, including the effects of galaxy mergers, but do not include spatial information \citep[except in a simplified form in more recent models,][]{henriques_l-galaxies_2020,stevens_dark_2024}.
These simulations have been used to forward model the galaxy population in large periodic volumes \citep{kitzbichler_high-redshift_2007,henriques_effect_2011,somerville_galaxy_2012}.
Cosmological hydrodynamic simulations explicitly model the 3D star-gas-black hole geometry, with detailed merger and assembly histories \citep{crain_hydrodynamical_2023}, and each stellar element (or particle) can be treated as an individual co-eval population.

Photoionisation modelling can be used to estimate the contribution of nebular line and continuum emission from young stellar populations still embedded in their birth clouds \citep{charlot_nebular_2001,gutkin_modelling_2016,byler_nebular_2017,wilkins_nebular-line_2020}. This is typically achieved using photoionisation codes such as \textsc{Cloudy} \citep{ferland_2017_2017,chatzikos_2023_2023} and \textsc{Mappings} \citep{dopita_spectral_1996,MappingsIII_2008}.
It can also be used to predict the emission from far-infrared (FIR) cooling lines in the diffuse interstellar medium \citep{Vallini_2015_CII,sigame_2017}, including molecular lines such as CO and HCN.
Many resonant lines require radiative transfer approaches to model accurately, such as Lyman-$\alpha$.
The energetic surroundings of AGN can also be modelled using photoionisation codes \cite[e.g.][]{feltre_nuclear_2016,hirschmann_synthetic_2017}, distinguished using the unified AGN model into the disc, torus, narrow line region (NLR) and broad line region (BLR).
These models typically assume simplified templates, often empirical, to reduce the dimensionality of the problem.

Most models are further distinguished by their treatment of dust; the simplest use idealised screen models, where the dust is represented by a slab of dust in front of the stellar distribution \citep{charlot_simple_2000,trayford_colours_2015,trayford_it_2016}.
The most sophisticated are three dimensional dust radiative transfer (RT) approaches \citep{steinacker_three-dimensional_2013}, typically employing Monte Carlo (MC) methods \citep{whitney_monte_2011}, which account for scattering and thermal re-emission.
A number of codes have been developed, such as \textsc{Skirt} \citep{camps_skirt_2015,skirt9}, \textsc{Sunrise} \citep{jonsson_sunrise_2006}, \textsc{ART}$^2$ \citep{li_art2_2020} and \textsc{Powderday} \citep{robitaille_hyperion_2011,narayanan_powderday_2021}.
These have been applied to cosmological hydrodynamic simulations \citep{torrey_synthetic_2015,trayford_optical_2017,camps_data_2018,lovell_reproducing_2021,baes_tng50-skirt_2024}, as well as isolated galaxy simulations \citep[][]{lanz_simulated_2014,Behrens_skirt_2018} and cosmological zoom simulations \citep{cochrane_predictions_2019,Liang2019,punyasheel_first_2025}.
The distribution of enriched gas can be linked to that of the dust, or in some cases explicit treatment of dust formation and evolution, that self consistently predict the mass, abundance and grain size distribution of the dust \citep{bekki_cosmic_2015,aoyama_galaxy_2017,mckinnon_simulating_2017,li_dust--gas_2019}.
Line-of-sight dust models are an alternative approach, that use the star dust geometry to calculate the column density of dust (or metallicity proxy) along a given line of sight to each star particle in a simulation \citep{wilkins_properties_2017,vijayan_first_2021}.
These methods are much cheaper than full radiative transfer approaches, whilst still capturing the 3D star-dust geometry; however, they do not include scattering effects, nor the thermal re-remission.

Finally, a number of codes exist for transforming the forward modelled outputs into data products that closely resemble those produced from observational instruments \citep{harborne_simspin_2023,gabrielpillai_espresso_2024}.
A number of codes wrap these various aspects into full pipelines, that can generate field-level observables \citep{fortuni_forecast_2023,marshall_forecastor_2025}.

These codes have been used to make ground-breaking advances in our understanding of the physics driving observed properties of galaxies.
However, we argue that current codes have a number of limitations.
Many require significant computational resources (such as RT codes), often comparable to those used to produce the parent numerical simulations in the first place.
As a result of this cost, it is often expensive to apply them to large cosmological populations of simulated galaxies.
It is also often not possible to explore the dependence of the forward model on certain modelling choices, both due to inflexibility in the specific implementation, as well as computational cost; each of these choices can have a significant impact on the emission produced, and introduce non-trivial qualitative uncertainties in the predictions \citep[e.g.][]{wilkins_lyman-continuum_2016,Trcka2022_tng50}.
Producing various observational data products, such as photometry, spectra, imaging and integral field unit data cubes in a consistent manner using the same modelling choices is also often not supported.
The same limitation applies when running models on different numerical simulations; usually, new pipelines are written for each galaxy evolution code, utilising different physical properties and definitions from within the code, which complicates fair comparisons between simulations.
Finally, combining models for the stellar, gas and AGN emission in a self-consistent manner is often not supported, requiring separate codes with inconsistent assumptions, particularly around the handling of dust attenuation. To the best of our knowledge, there is no publicly available software or package that provides all the functionality described above.

In addition, new accelerated methods utilising neural density estimators, known as Simulation Based Inference (SBI) methods, have emerged in recent years \citep{cranmer_frontier_2020,ho_ltu-ili_2024}.
These methods require large synthetic training sets, and can easily accommodate wide sets of parameter variations, to test degeneracies in inference schemes.
Generating such large training sets requires rapid forward models of the integrated and resolved galaxy emission, as well as field-level instrumental effects \citep{lovell_learning_2024,fischbacher_ufig_2024,fischbacher_galsbi_2025,cakir_fast_2024}.
Population level models, such as pop-cosmos \citep{alsing_pop-cosmos_2024}, also require rapid forward models in order to train emulators \citep{alsing_speculator:_2020}, allowing hierarchical inference throughout the forward model.

\textsc{Synthesizer}\footnote{\url{https://synthesizer-project.github.io}} addresses many of these drawbacks in current forward modelling codes, as well as targeting opportunities in emerging novel inference schemes.
The main design principles we adopted when developing \textsc{Synthesizer} are as follows: 
\begin{itemize}
    \item \textbf{Flexible}: where something can be changed, it should be changeable.
    \item \textbf{Fast}: speed should be prioritised over fidelity, and where it does not sacrifice simplicity
    \item \textbf{Modular}: elements of the code should be reusable and multi-purpose.
    \item \textbf{Extensible}: new capabilities and functionality should be easy to adopt and integrate.
\end{itemize}
The flexibility of the code ensures that many of the uncertain parameters in forward modelling (or SED fitting) can be modified to assess their impact on the results. 
The code is intended to be fast to enable a thorough exploration of uncertain model parameters on large simulation catalogues.
For example, the code is intended to be fast enough that multiple different SPS and dust models can be run on the same set of simulated galaxies to explore the dependence of their emission on these qualitative modelling assumptions.
The modularity of the code ensures that the same elements and objects can be re-used for multiple tasks.
At the highest level, this means that much of the code is usable for both forward and inverse modelling, as well as for generating photometry, spectra, imaging and integral field unit data products. 
Finally extensibility ensures that any new functionality, such as the adoption of accelerated forward models, can be easily accommodated within the \textsc{Synthesizer} ecosystem, enabling easy integration with other modules and functionality.
In support of the above principles, the code is open source, version controlled, has detailed integration test coverage, as well as coherent and up to date documentation.
See \cite{roper_synthesizer_2025} for further details on the code implementation.

\textsc{Synthesizer} is not intended to replace high fidelity dust or line-radiative transfer approaches.
It is instead intended to be a platform for performing rapid predictions, that can help users understand the impact of certain choices in their forward model.
In fact, \textsc{Synthesizer} can be thought of as complementary to dust RT codes, and can also interface with them through some of the higher level data processing modules.

We provide an overview of the software design in \sec{design}, with links to various sections describing the main components.
We present our conclusions and a summary in \sec{conc}.
More detailed descriptions of particular functionality is provided in the appendices, and linked where appropriate throughout the text.

\section{Design Overview}
\label{sec:design}

\begin{figure*}
	\includegraphics[width=\textwidth]{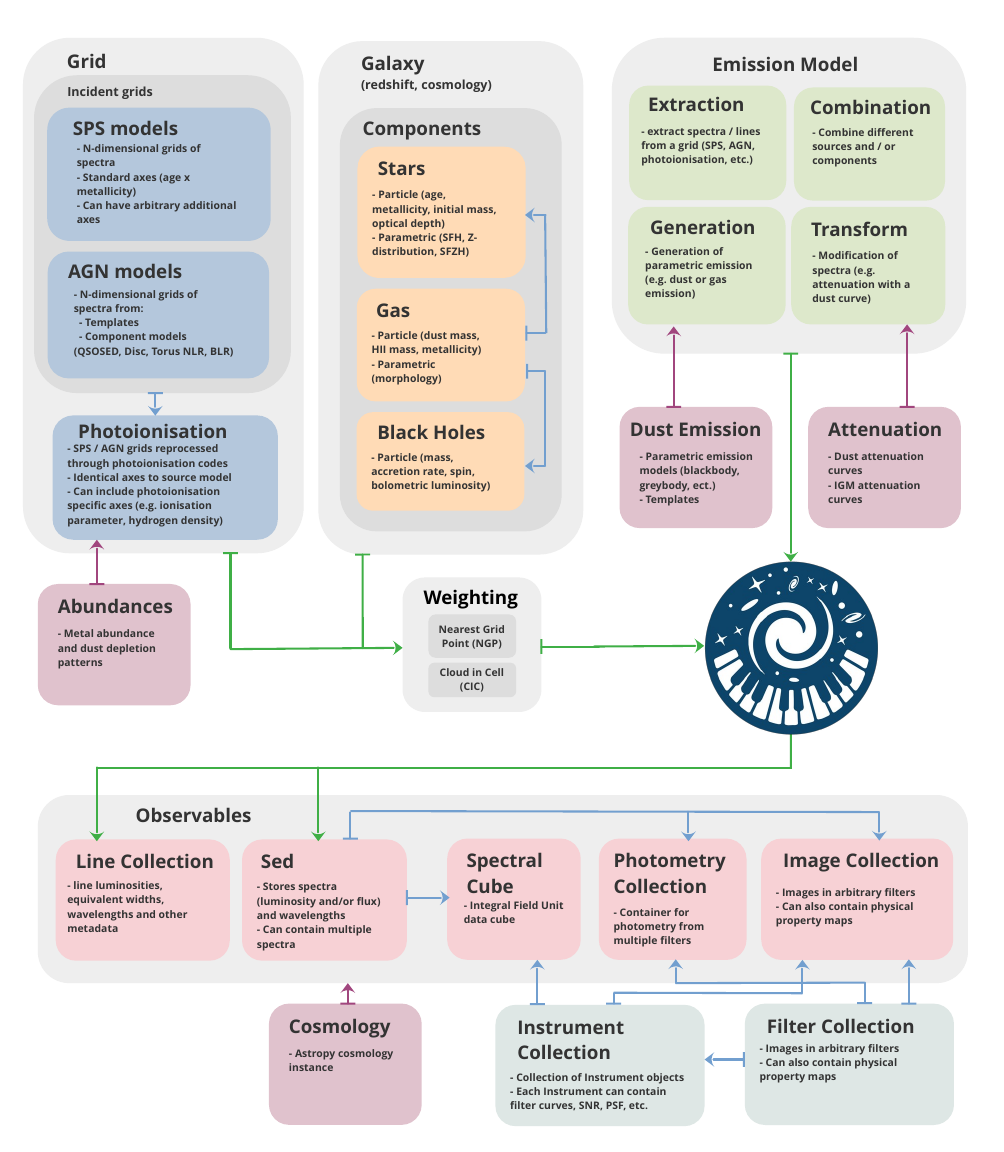}
    \caption{Schematic showing the main modules available in \textsc{Synthesizer} and how they relate to each other, described in \sec{design}. Most forward modelling applications will combine a Grid object, a Galaxy object (consisting of stars, gas and black holes components) and an Emission Model object, and produce a number of observables (Sed's, images, etc.).}
    \label{fig:modules}
\end{figure*}

\textsc{Synthesizer} consists of a number of interlinked modules, shown in \fig{modules}.
In this section we provide a high level overview of each component and how they relate to each other, before describing them in greater detail in later sections as part of the illustrative examples.

The main component in \textsc{Synthesizer} is the \textsc{Galaxy} object.
This is a wrapper for all of the various physical components that a galaxy contains, such as stars, gas and black holes. 
\textsc{Galaxy} objects are split into two main types, \textit{parametric} or \textit{particle}.
The former contains descriptions of the star formation and metal enrichment history as described by either individual or arbitrary combinations of parametric functional forms.
\textit{Particle} galaxy objects instead contain components described by collections of particles, e.g. from a smoothed particle hydrodynamics simulation.
Both \textit{particle} and \textit{parametric} \textsc{galaxy} components share very similar functionality, but differ in their lower level implementations.
Further details on \textsc{galaxy} components are provided in \sec{galaxy}.

Another key component in \textsc{Synthesizer} is the \textsc{Grid} component.
Grids in \textsc{Synthesizer} parlance are $N$-dimensional arrays of the underlying emission models, that are then coupled to various physical galaxy properties to produce the emission for a particular object.
The simplest and most commonly used grid is a pure SPS model, describing the emission at a range of wavelengths along age and stellar metallicity dimensions. 
However, grids can take arbitrary numbers of dimensions, and these can describe arbitrary physical parameters.
The same grid structure is used to describe SPS models processed through photoionisation codes, as well as the emission from AGN.
Further details on \textsc{Grid} components are provided in \sec{grids}.

Grid and galaxy components can be combined to produce synthetic observables using \textsc{Emission Models} (see \sec{emission_models}).
These describe how the properties of a \textsc{Galaxy}, such as its star formation and metal enrichment history, are coupled to a \textsc{Grid(s)}, as well as additional observable effects such as dust attenuation or emission.
Dust attenuation models represent a range of sophistication, from simple screen models to line-of-sight models that account for the full star--dust geometry.

\begin{figure*}
	\includegraphics[width=\textwidth]{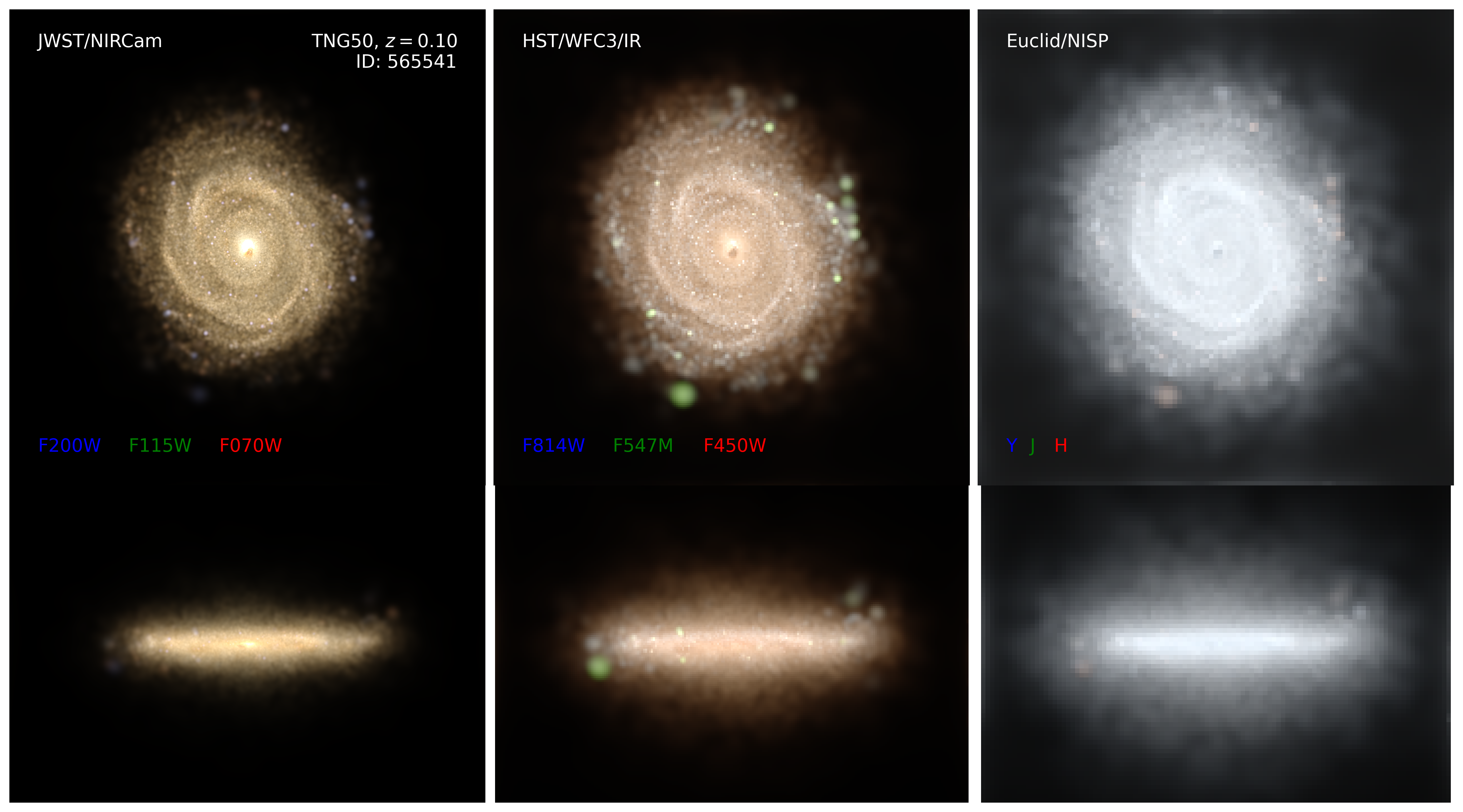}
    \includegraphics[width=\textwidth]{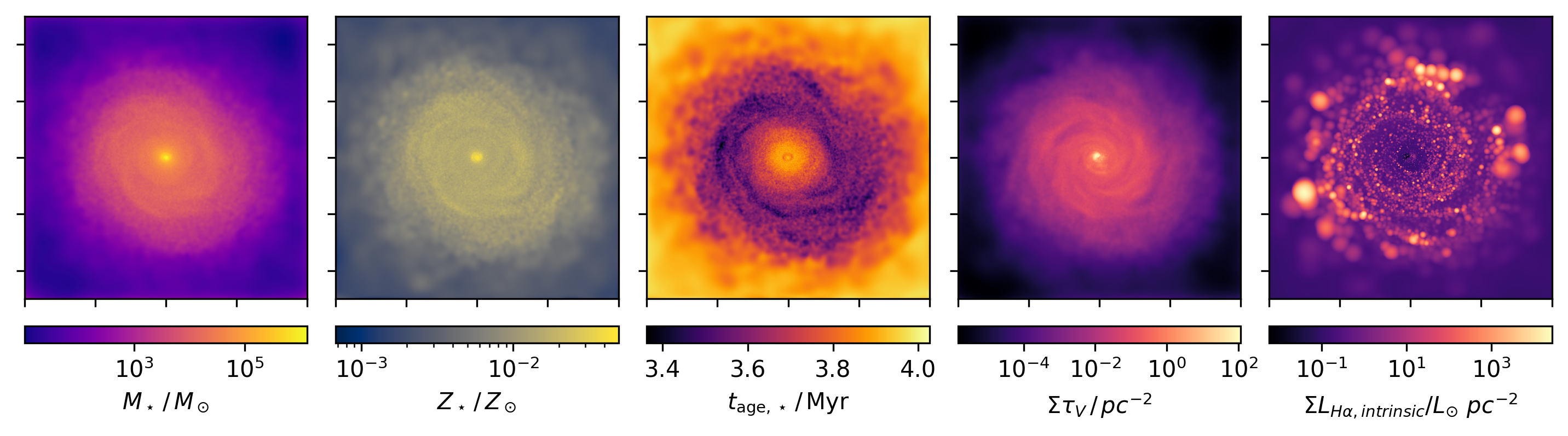}
    \includegraphics[width=\textwidth]{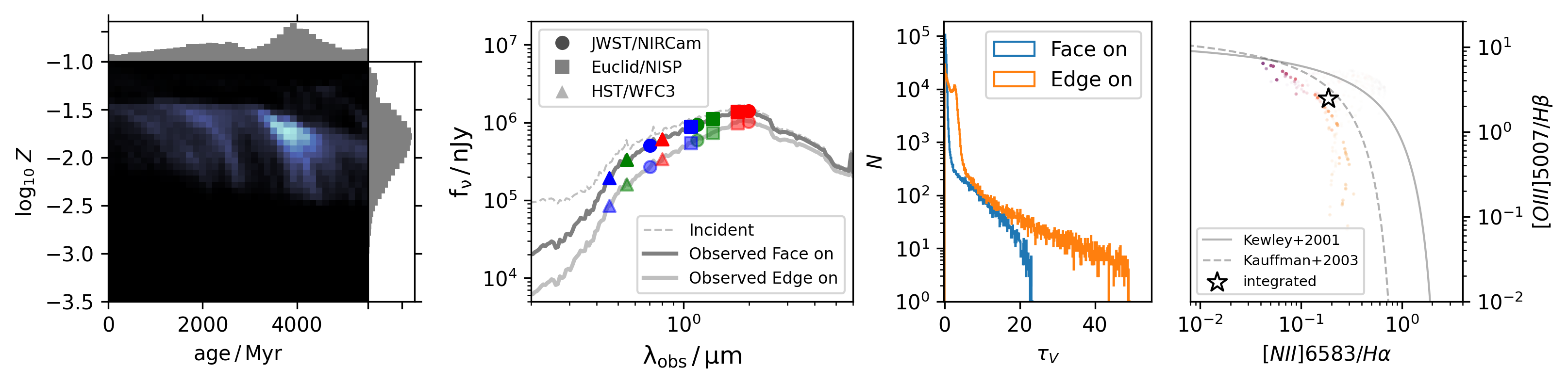}
    \caption{An example showing some of the forward modelled observables that can be produced using \textsc{Synthesizer}. We show a galaxy taken from the Illustris TNG50 simulation, processed using a line of sight dust attenuation model, assuming a fixed dust to metal ratio, and using the BPASS v2.2.1 SPS model, with binary stars and a Kroupa IMF. \textit{Top row}: full colour RGB images in, from left to right, JWST/NIRCam [F115W,F150W,F200W], HST/WFC3 [F450W,F547M,F814W], and Euclid NISP [YJH], where each band is linearly combined. The NIRCam image is produced using a PSF taken from \cite{perrin_updated_2014}. \textit{Middle row, left to right}: stellar mass map, mass-normalised metallicity map, mass-normalised age map, optical depth map (obtained from the gas particles), H-$\alpha$ emission map. \textit{Bottom row, left to right}: the star formation - metal enrichment history (with marginal age and metallicity distributions), intrinsic and dust attenuated spectra and derived photometry (for edge on and face on orientations), histogram of optical depth for face-on and edge-on line of sight calculations, BPT diagram (star shows the integrated ratios, points show individual star particles, lines show the \protect\cite{kauffmann_unified_2000,kewley_theoretical_2013} classification regions).
    }
    \label{fig:mosaic}
\end{figure*}

\begin{figure*}
    \centering
    \includegraphics[width=1.0\textwidth]{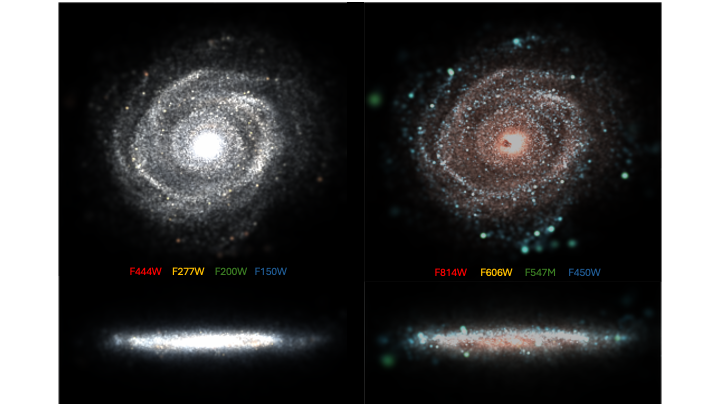}
    \caption{An idealised ROGB image of the same galaxy as in Figure~\ref{fig:mosaic} in the JWST NIRCam F444W, F277W, F200W, F150W (left) and HST WFPC2 F814W, F606W, F547M, F450W (right) filters, using the \texttt{make\_lupton\_rgb} scheme in \texttt{astropy} using custom weightings. The top panel shows the face-on view while the bottom panel shows the edge-on view. This utilises a higher angular resolution compared to the native resolution of JWST and HST.}
    \label{fig:idealised_rogb}
\end{figure*}

Once the emission has been generated it can be presented or transformed into a variety of different observable formats (see Sections \ref{sec:generating_spectra} and \ref{sec:emissions}).
These include high resolution SEDs, represented by \textsc{Sed} objects, RGB images, integral field unit spectral cubes, emission lines and photometry.
The latter two can be combined into Line and Photometry collections.
Some of these outputs are shown in \fig{mosaic} for an example galaxy from the IllustrisTNG simulation suite. We also present an idealised ROGB image of the same galaxy in \textit{JWST} and \textit{HST} filters in Figure~\ref{fig:idealised_rogb}, using custom weightings for the different filters.

Finally, we describe the \textsc{Pipeline} module (see \sec{pipeline}), which provides a single interface to all of the above functionality, and streamlines the implementation of OpenMP and MPI parallelism (described in \sec{parallelism_scaling}).

\textsc{Synthesizer} implements its own unit system (Appendix~\ref{sec:units}).
Further details on other advanced functionality, such as the ability to define custom abundance patterns, are provided in the extensive appendices.

\vspace{1.0cm}

\section{Galaxy Objects and Components}
\label{sec:galaxy}

A \texttt{Galaxy} object is essentially a container object for different components (stars, gas, and black holes) that can make up a galaxy, and provides methods for interacting with and combining these components.
In addition to the component attributes, this can also hold galaxy level attributes, such as redshift (required to calculate the observer frame emission).

\texttt{Galaxy} objects can take two different forms depending on the underlying data representation: particle or parametric.
A Parametric \texttt{Galaxy} describes the star formation and metal enrichment (SFZH) history, as well as the 3D morphology, through combinations of parametric forms.
Conversely, a Particle \texttt{Galaxy} object is comprised of individual star, gas, and/or black hole particles, usually derived from a hydrodynamic simulation\footnote{We will refer to this form as Particle, but this can be any choice of the resolution element in the simulation whose extent can be defined through a smoothing kernel}.

\texttt{Galaxy} objects in \textsc{Synthesizer} consist of a number of different components.
These can include stellar, gas or black hole components.
Each of these components can be initialised independently of a Galaxy object, and each can be coupled with a Grid object to produce their own emission.
They can also take particle or parametric forms in most cases.
Below we provide an overview of each.

\subsection{Stellar Components}
\label{sec:stars}

\begin{figure*}
	\includegraphics[width=\textwidth]{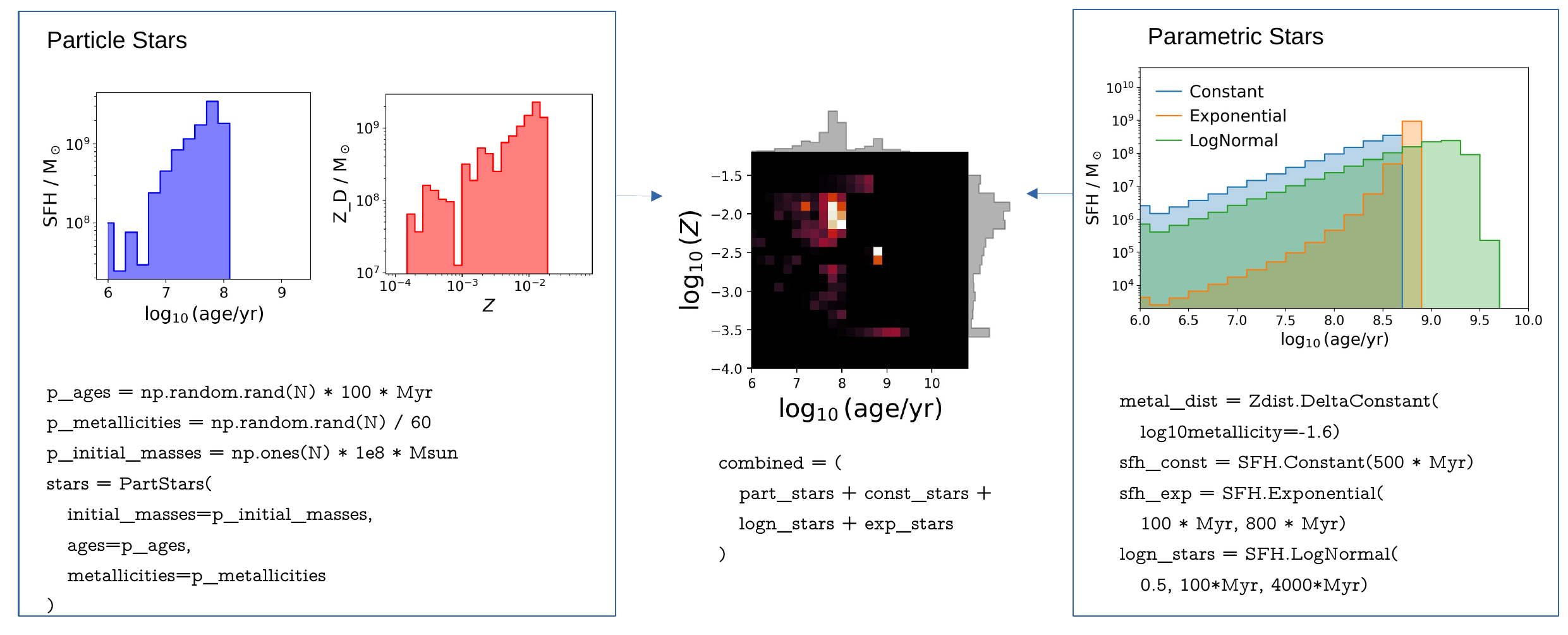}
    \caption{Example of defining parametric and particle star formation and metal enrichment histories (SFZH). \textit{Left panel}: randomly initialised star particles with a range of ages and metallicities, and their marginal distributions. \textit{Right panel}: three parametric star formation history forms provided in \textsc{Synthesizer}, a constant, exponential and log-normal star formation history (SFH), each at fixed metallicity. \textit{Middle panel}: The composite star formation and metal enrichment history (SFZH) from the combination of all the above particle and parametric objects. This functionality allows the generation of arbitrarily complex SFZH distributions.}
    \label{fig:sfh_intro}
\end{figure*}

A particle stars component requires at minimum arrays describing the ages, metallicities and initial masses of the particles.
Together these describe the SFZH of the galaxy (see \fig{sfh_intro}).
Additional optional information can be provided.
This includes the coordinates, which are necessary for imaging, LoS (Line-of-Sight) dust attenuation, and for filtering particles based on 3D or projected apertures.
Velocities can also be provided, which are used for line broadening.

A parametric stars object defines a 2D SFZH, along with a value for the total initial stellar mass.
There are a number of ways of describing this SFZH grid.
Users can explicitly provide a 2D array, which will then be converted to the underlying SPS grid dimensions if a grid is provided.
Individual 1D distributions can be provided, which will necessarily not include any covariances between the age and metallicity.
If an instantaneous burst is desired, scalars for the age and metallicity can be provided.
Once defined, this SFZH grid can be coupled with an SPS grid to predict the emission.
Finally, distribution functions for the star formation history and metal distribution can be defined.
We provide a number of pre-defined forms, including for the SFH a Constant, Gaussian, Exponential, Truncated Exponential, Declining Exponential, Delayed Exponential plus Log-normal, and a Double Power Law (some of these are shown in \fig{sfh_intro}), as well as the Dense Basis prescription \citep{iyer_reconstruction_2017,iyer_nonparametric_2019}.
For the metal distribution we provide a Delta Constant (scalar), or Normal distribution. 
Users can also define their own forms using a standardised class format.
Some of these forms are shown in \fig{sfh_intro}.

Both particle and parametric components can be combined to form arbitrarily complex SFZH distributions.
Parametric stellar \textit{morphologies} can also be described, detailed in \app{parametric_stellar_morph}.

\subsection{Gas Components}
\label{sec:gas}

A Gas component describes the gaseous medium in a galaxy.
\textsc{Synthesizer} currently only implements a Particle gas component; we do not define a parametric Gas component in the current version.\footnote{This will be implemented in a future version where the gas emission more generally is included.}

A Gas component requires at minimum the particle masses and metallicities.
Gas components can also additionally contain information such as the coordinates, smoothing lengths, dust masses and / or the dust to metal (DtM) ratios, necessary for calculating the LoS attenuation.
If the DtM ratio is not provided then a default $\rm DtM = 0.3$ is assumed \citep{camps_far-infrared_2016,schulz_redshift-dependent_2020}.
If dust masses are not provided then this DtM is used to automatically calculate the dust masses.
Other `subgrid' properties of gas particles can also be defined, such as the HI and HII mass fractions.

\subsection{Black Hole Components}
\label{sec:black_holes}

Unlike for stars the division between a particle and a parametric \texttt{BlackHoles} / \texttt{BlackHole} component is not well defined.
If you are only interested in exploring an AGN emission model without coupling to simulation data directly, you can use a \texttt{parametric.BlackHole} object.
A parametric \texttt{BlackHole} object has some specific differences compared to a \texttt{particle.BlackHole} object:
\begin{itemize}
    \item A \texttt{parametric.BlackHole} can only ever describe a singular black hole.
    \item A \texttt{parametric.BlackHole}'s “position” (i.e. if making an image) is described by a PointSource morphology object rather than coordinates.
    \item A \texttt{parametric.BlackHole} exists in isolation, i.e. it does not interface directly with other parametric components.
\end{itemize}

A particle \texttt{BlackHoles} object, conversely, allows the grouping of multiple black holes, each with their own coordinates, and can interface with other particle and parametric components.
\texttt{BlackHoles} objects can be initialised with arrays describing their masses, coordinates and accretion rates.
These are the minimum parameters required to enable the computation of the emission from an AGN grid.\footnote{Note that masses and accretion rates are positional arguments, and must therefore always be provided for \texttt{particle.BlackHoles}, while \texttt{parametric.BlackHoles} have more flexibility.}
Additional parameters can be provided, such as the inclination (relative to the observers line of sight) and environmental metallicities.
If masses and accretion rates ($\dot{M}_{\bullet}$) are provided, bolometric luminosities, $L_{\rm \bullet,bol}$, will be calculated automatically,
\begin{align}
    L_{\rm \bullet,bol} = \epsilon_r \dot{M}_{\bullet} c^2 \;,
\end{align}
where $\epsilon_r$ is the radiative efficiency, which defaults to 0.1 but can be specified explicitly for each black hole, and $c$ is the speed of light.
The Eddington ratio, Eddington accretion rate and Eddington luminosity are also calculated.

Finally, the composition of the gas surrounding a black hole can have a significant impact on its emission, controlling the level of dust and nebular attenuation, and the form of the re-emitted radiation.
As such, \textsc{Synthesizer} allows the user to specify the metallicity of the surrounding gas.
This can also be computed from the properties of nearby gas particles where this information is present, providing a more physically-motivated value, though we note that the scales on which this is typically computed for cosmological simulations is many orders of magnitude larger than the accretion disk around the black hole. 
Further details on calculating the emission from black holes is provided in \sec{emission_models}.

\section{Grids}
\label{sec:grids}

\begin{figure*}
	\includegraphics[width=\textwidth]{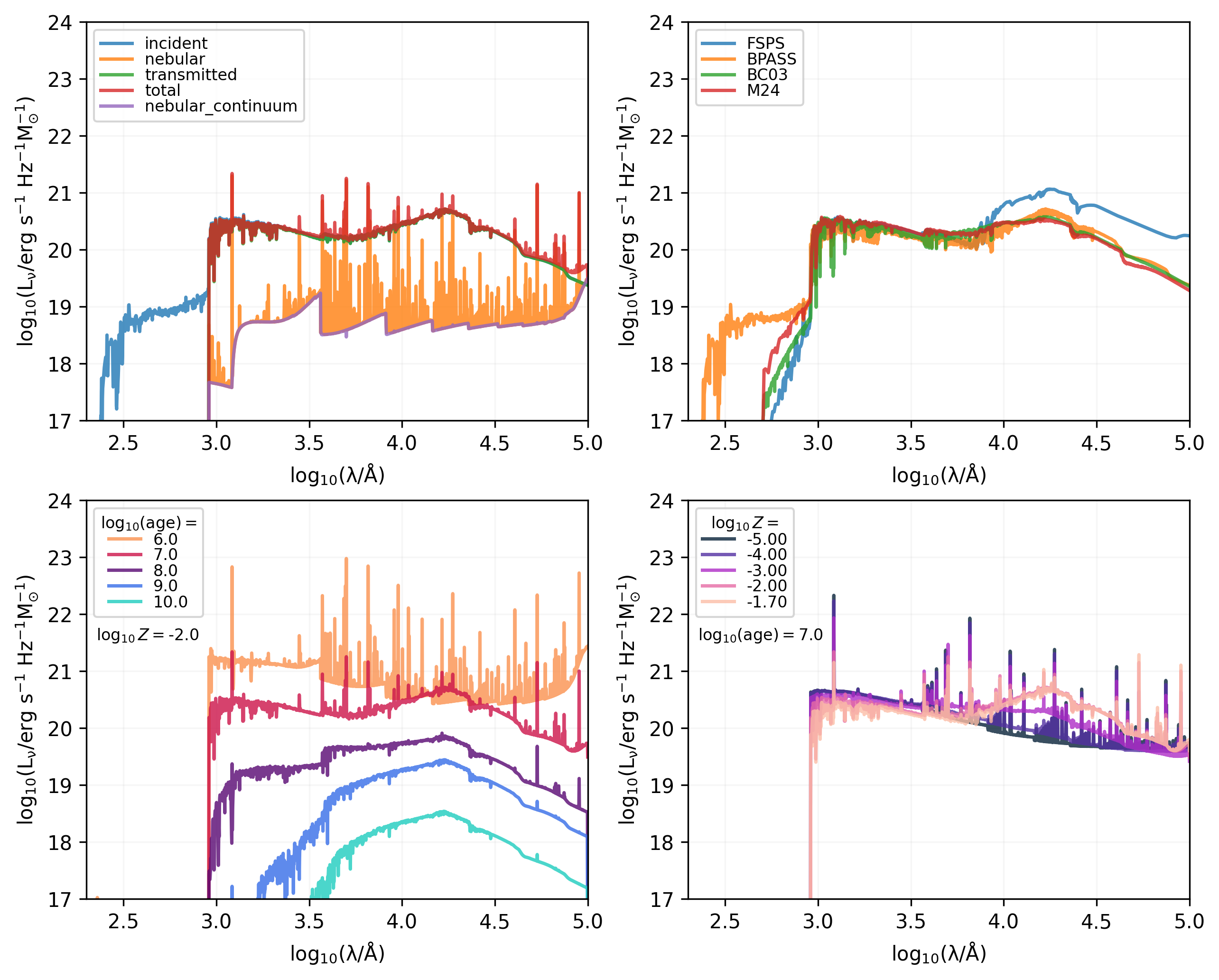}
    \caption{Example of the spectra contained within a \textsc{grid} object. \textbf{Top left}: all the different spectra produced from a grid, with age 5 Myr and metallicity 0.01, using the BPASS v2.2.1 SPS binary model with a Kroupa IMF, and post-processed with Cloudy with the fiducial photoionisation parameters. \textbf{Top right}: multiple incident grids assuming different SPS models (BC03, BPASS, FSPS, M24) all at the same age (5 Myr) and metallicity ($Z = 0.01$) \textbf{Bottom left}: variation of the same model in the top left panel with age, at fixed metallicity ($Z = 0.01$). \textbf{Bottom right}: variation of the same model in the top left panel with metallicity, at fixed age (5 Myr).}
    \label{fig:grid_example}
\end{figure*}

\texttt{Grids} are a fundamental concept in \textsc{Synthesizer}, and almost all other functionality is reliant on them.
They are typically precomputed multi-dimensional arrays of observables, such as spectra and lines, stored in HDF5 files that can be loaded at run time. 
Grids can represent the emission from a range of sources, commonly stellar populations (as computed through Stellar Population Synthesis (SPS) models), but also Active Galactic Nuclei (AGN) and diffuse interstellar gas.

We provide a number of pre-computed grids, but users can also generate their own grids using the \texttt{grids} package\footnote{\url{https://github.com/synthesizer-project/grid-generation}} (see \app{grids_create} for details).
Below we provide more details on the main types of \texttt{Grid}. 

\subsection{Stellar Population Synthesis Grids}
\label{sec:sps_grids}

Grids based on Stellar Population Synthesis (SPS) models all have at least two dimensions, age and metallicity.
For each grid point we provide the full spectrum, where the wavelength array is identical for all grid points.
A number of derived properties are computed on a grid, such as the specific ionising luminosity (where there is sufficient coverage blueward of the lyman-break).

The naming of SPS grids in general follows this specification:
\begin{lstlisting}
{sps_model}-{sps_version}-{sps_variant}_
{imf_type}-{mass_boundaries}-{slopes}_
{photoionisation_code}-
{photoionisation_code_version}-
{photoionisation_parameters}
\end{lstlisting}
e.g.
\begin{lstlisting}
bpass-2.2.1-bin_chabrier03-0.1,300.0_
cloudy-c17.03
\end{lstlisting}
specifies that the grid is constructed using v2.2.1 of the Binary Population and Spectral Synthesis (BPASS) SPS model for the binary (bin) variant, assumes a \cite{chabrier_galactic_2003} IMF between 0.1 and 300 $M_{\odot}$, and that photoionisation modelling is performed using v17.03 of the \textsc{Cloudy} photoionisation code assuming our default assumptions (more details below).
Many grids will omit parameters in the name where they take the default values.
It is also worth noting that all grids files contain attributes describing the parameters used in detail, with units, and these should be the values used when referencing programmatically.

Grids are constructed using various initial mass functions (IMFs), depending on their availability in the original SPS model.
The \cite{chabrier_galactic_2003} IMF is available in most SPS models, allowing a like-for-like comparison.
To explore the systematic impact of changing the IMF, broken power law (bpl) IMFs are more suitable. These are named as so, \verb|{imf_type}-{mass_boundaries}-{slopes}|. For a Salpeter (1955) IMF (slope=2.35) between 0.1 and 100 Msol, \verb|bpl-0.1,100-2.35|.
A more complex IMF, for example with two power-laws (2.0, 2.35) separated at 1 Msol, \verb|bpl-0.1,1.0,100-2.0,2.35|.

\tab{sps_grids} lists details of some of the main SPS models provided with \textsc{Synthesizer}.
A number of example stellar emission grids are shown in \fig{grid_example}, including the various spectral components, and variation with age and metallicity.
A number of different grids from different SPS models are also shown, at a similar fixed age and metallicity, highlighting the differences in the UV-optical emission.

\begin{table*}
	\centering
	\caption{Incident stellar population synthesis grids}
	\label{tab:sps_grids}
	\begin{tabular}{lccccc}
		\hline
		Name & Metallicities & Ages / yr & Wavelengths / \textup{~\AA} & Isochrones & Reference \\
		\hline
		BC03 & 0.0004 - 0.03 (6) & $10^5 - 10^{10}$ (221) & $10^2 - 10^6$ (6900) & Padova + Geneva & \protect\cite{bc03} \\
		BPASS & 0.00001 - 0.04 (13) & $10^6 - 10^{11}$ (51) & $10^0 - 10^5$ (100,000) & BPASS isocontours & \protect\cite{bpass} \\
		FSPS & 0.00004 - 0.04 (12) & $10^5 - 10^{10}$ (107) & $10^2 - 10^8$ (1963) & MIST & \begin{tabular}{@{}c@{}}\protect\cite{fsps1}; \\ \protect\cite{fsps2}\end{tabular}  \\
        Maraston & 0.001 - 0.04 (4) & $10^3 - 10^{10}$ (67) & $10^2 - 10^6$ (1221) & Geneva & \begin{tabular}{@{}c@{}}\protect\cite{maraston_evolutionary_2005}; \\ \protect\cite{newman25}\end{tabular} \\
        Yggdrasil & NA & $10^4 - 10^{9.5}$ (111) & $10^2 - 10^8$ (1221) & Padova & \protect\cite{Yggdrasil2011} \\
		\hline
	\end{tabular}
    \tablecomments{Further details on each model are provided in the listed references. For the isochrones, Padova refers to the collection of isochrones by \protect\cite{alongi93,bressan93,fagotto94a,fagotto94b,girardi96,girardi00} and the Geneva isochrones refer to the compilation by \protect\cite{schaller92,charbonnel96,charbonnel99}. For information on the MIST and BPASS isocontours, see \protect\cite{dotter16,choi16} and Section 2 of \protect\cite{bpass}, respectively.
    Yggdrasil is an SPS model specifically designed for Population III stars, and as such does not have a typical grid metallicity structure due to the predominance of metal free models.}
\end{table*}

\subsection{Active Galactic Nuclei Grids}
\label{sec:grids_agn}

\textsc{Synthesizer} includes the ability to model the emission from accreting super-massive black holes (SMBH), i.e. active galactic nuclei (AGN). In keeping with the \textsc{Synthesizer} philosophy, multiple approaches are available. In the most basic approach we have a single - typically empirical - template that is simply scaled by the bolometric luminosity of the AGN. Although no specific grids are currently available, a simple extension to the single template described above is a grid of templates, for example for different SMBH masses or bolometric luminosities. In this case \textsc{Synthesizer} will automatically associate the closest template to the AGN being modelled. 

We have also developed a more sophisticated model, termed \emph{UnifiedAGN}.
This model, which will be fully described and explored in Wilkins et al. \emph{in-prep}, but is briefly described in \S\ref{sec:em_agn}, self-consistently combines emission from the accretion disc, line emitting regions, and a dusty torus. Supporting this model is a suite of grids including photoionisation modelling encompassing the physical conditions expected in both narrow and broad line emitting regions. These grids are built using a variety of different disc models, currently including a simple broken power-law parameterisation \cite[similar to that in][]{feltre_nuclear_2016}, and the physically motivated \texttt{qsosed} \citep{kubota_physical_2018} and \texttt{relqso} \citep{hagen_estimating_2023} models. \texttt{qsosed} is a simplification of the more general \texttt{agnsed} model where the only free parameters are the SMBH mass and accretion rate. Similarly, \texttt{relqso} is a simplification of \texttt{relagn} which extends the \texttt{agnsed} model to include relativistic effects, adding SMBH spin as an additional free parameter. For each of these disc models we create a grid of intrinsic spectra, varying the free parameters. 

The naming convention for AGN grids follows a structure similar to that used for SPS grids, combining information about the disc “incident” model and the photoionisation setup. However, a key distinction is that, unlike SPS models — whose grids are fixed by the underlying stellar population models — we have direct control over the parameter space sampled by AGN disc models. For instance, using the \texttt{qsosed} model, we can generate arbitrarily fine grids in SMBH mass and accretion rate. While generating these disc models is computationally inexpensive, the addition of photoionisation modelling can be costly, especially with multiple free parameters. To balance flexibility and efficiency, we therefore provide multiple variants at different grid resolutions.
\texttt{qsosed} also allows us to specify the viewing angle of the emission. In some applications, however (such as modelling nebular line and continuum emission) it is the inclination-averaged emission that is more physically appropriate. For this reason, we also provide versions of the models that are averaged over viewing angle.

The naming of AGN grids in general follows this specification:
\begin{lstlisting}
{disc_model}-{disc_model_version}-
{disc_grid_parameters}_
{photoionisation_code}-
{photoionisation_code_version}-
{photoionisation_parameters}
\end{lstlisting}
e.g.
\begin{lstlisting}
qsosed-isotropic_
cloudy-c23.01-blr
\end{lstlisting}

\subsection{Photoionisation Processed Grids}

In order to model the effects of photoionisation of astrophysical gas, source grids can be processed through photoionisation codes to predict the line and continuum emission from that gas.
The sources of the ionising photons can be stellar populations or AGN, and the gas itself can represent a dense nebular cloud, diffuse interstellar gas, or gas surrounding an AGN.
In all these cases, we post-process each source grid with a photoionisation code under different assumptions for the gas; this processed grid can then be used in concert with other source grids to model the composite emission from various distinct components within a galaxy.
Below we describe our fiducial approach for photoionisation modelling in synthesizer, with a focus on the nebular emission from young stellar populations, though we note that other approaches can be adopted and used to produce grids as defined by the user, for arbitrary source and gas configurations. 
The photoionisation modelling of the AGN grids will be fully described in Vijayan et al. \emph{in-prep}.
We use the \textsc{Cloudy} photoionisation code \citep{ferland_2017_2017,chatzikos_2023_2023} for all fiducial grids in synthesizer, though other codes can be substituted.

\subsubsection{SPS Grids}
\textsc{Synthesizer} adopts a similar method for modelling the nebular emission from stellar sources to \cite{wilkins_nebular-line_2020,vijayan_first_2021}; both these studies follow the approach of \cite{charlot_nebular_2001, gutkin_modelling_2016, feltre_nuclear_2016}, though with some key differences.
The ionising source, in this case a simple stellar population (SSP), is treated as a point source, surrounded by concentric spherical layers of gas.
The intensity of the ionising spectrum ($\lambda_0 \leqslant 912 \; \mbox{\normalfont\AA}$) from the SSP is given by
\begin{align}
    Q = \frac{1}{hc} \int^{\lambda_0}_0 \lambda f_{\lambda} d\lambda \;\;,
\end{align}
where $Q$ is the number of ionising photons capable of ionising hydrogen.

The \textit{ionisation parameter}, $U$, is a useful dimensionless parameter that represents the ratio of ionising photons to gas densities at a distance $R$ from the source,
\begin{align}
    U(R) = \frac{Q}{4 \pi R^2 n_{\mathrm{\mathrm{H}}} c} \;\;.
\end{align}
where $c$ is the speed of light, and $n_{\mathrm{H}}$ is the hydrogen density.
It is derived at the Str{\"o}mgren radius $R_{\mathrm{S}}$, where the rates of ionisation and recombination for a pure hydrogen nebula are in thermal equilibrium. $R_{\mathrm{S}}$ is defined as
\begin{align}
    R_S^3 = \frac{3Q}{4 \pi n_{\mathrm{H}}^2 \epsilon \alpha_{\mathrm{B}}}
\end{align}
where $\epsilon$ is the volume-filling factor of the gas, and $\alpha_{\mathrm{B}}$ is the case-B hydrogen recombination coefficient \citep{osterbrock_astrophysics_1989}.
$U(R_{\mathrm{S}})$ can only be calculated after a photoionisation model has been run \citep{byler_nebular_2017}.
In \textsc{Cloudy}, the strength of the ionising source can be provided in terms of the ionisation parameter, but defined at the inner edge of the cloud, $R_{\mathrm{inner}}$.
An alternative parametrisation is that of the \textit{volume averaged} ionisation parameter, $\langle U \rangle$, which gives the average of the ionisation parameter throughout the ionised region.
The geometry of the ionised region is primarily governed by the ratio between the inner radius of the the cloud and the thickness of the cloud ($\Delta R$).
When $R_{\mathrm{inner}} \geqslant \Delta R$, the ionised region is thin, and the ionisation parameter remains approximately constant, producing essentially a plane parallel geometry.
In contrast, when $R_{\mathrm{inner}} << \Delta R$, the ionisation parameter varies significantly with $R$, leading to a spherical geometry.
In case of the SPS grids, we adopt a spherical geometry, for which the volume averaged ionisation parameter is given by,
\begin{align}
    \langle U \rangle \approx \frac{3 Q}{4 \pi R^2 n_{\mathrm{H}} c} &= 3 U(R_{\mathrm{S}}) \\
    &= \frac{\alpha_B^{2/3}}{c} \left( \frac{3Q \epsilon^2 n_H}{4 \pi} \right) \;\;.
\end{align}
For fixed values of $\langle U \rangle$ and $Q$, the geometry of the region is encapsulated in the $\epsilon^2 n_{\mathrm{H}}$ term.
The advantage of using $\langle U \rangle$ is that it significantly reduces the dimensionality of our input grid: any combination of $Q$, $\epsilon^2$ and $n_{\mathrm{H}}$ that yields the same $\langle U \rangle$ will result in an identical nebular spectrum -- provided the shape of the ionising spectrum and metallicity are held constant.

The synthesizer grids package allows users to vary the hydrogen density ($n_{\mathrm{H}}$) and inner radius $R_{\mathrm{inner}}$ in order to evaluate the effect of these assumptions on the derived observables (see \sec{grids_create}).
However, for our fiducial grids we assume a fixed geometry, with $R_{\mathrm{inner}} = 0.01 \; \mathrm{pc}$ and $n_{\mathrm{H}} = 10^{2.5} \; \mathrm{cm^{-3}}$; the only values that we choose to fix or vary are $\langle U \rangle$ and $Q$.

We provide grids for two approaches.
In the first, we use a fixed $\langle U \rangle$, and vary $Q$ to achieve this.
Since the value of $Q$ measured from the SSP's for a $1 \mathrm{M_{\odot}}$ population is often too small to achieve the desired ionisation parameter, it is necessary to normalise $Q$ for each SSP, as demonstrated in \cite{byler_nebular_2017}, effectively increasing the mass of the ionising source.
This normalising mass ($\hat{M}$) can then be used to renormalise the output of the photoionisation modelling for a $1 \mathrm{M_{\odot}}$ source population.
We explore a range of values, $\mathrm{log_{10}} \langle U \rangle = [-3,-2,-1]$.

In the second approach we allow $\langle U \rangle$ to vary with the input ionising source, but normalised to some reference value.
We again normalise the mass for this reference value, but use the same normalisation for all input models ($\hat{M}_{\mathrm{ref}}$).
For our fiducial grids, we choose to normalise to a value of $\mathrm{log_{10}} \langle U \rangle = -2$ for an age $1 \; \mathrm{Myr}$ source population with metallicity $\mathrm{log_{10}}(Z) = -2$.
We process all the fiducial grids presented in \tab{sps_grids} through \textsc{Cloudy} (with the default \textsc{Cloudy} energy mesh resolution). 

\fig{grid_example} shows the nebular line and continuum contribution to the total spectra of an age 5 Myr, metallicity 0.01 SSP from the BC03 model, assuming a Chabrier IMF.


\section{Emission Models}
\label{sec:emission_models}

In the simplest sense, generating the emission from a galaxy (i.e. spectra or emission lines) requires combining an emitter (i.e. a parametric or particle stars component, see Section \ref{sec:galaxy}) with a grid (see Section \ref{sec:grids}) and some prescriptions for how this light is affected by other components (e.g. dust).
In reality, there's a wealth of complexity that can be introduced.
A galaxy could include emission from different sources (e.g. stellar components and AGN), a single component could contribute multiple different emissions (e.g. nebular emission from young stellar populations, incident emission from older stars), or different components could experience different levels of dust attenuation, to name only a few sources of model complexity.
This complexity is compounded when the needs of specific astrophysical simulations are considered in conjunction with all these options.
For instance, some simulations self-consistently model the properties and distribution of dust \citep[e.g.][]{dave_simba:_2019,li_dust--gas_2019}, while others might include non-equilibrium chemistry \citep[e.g.][]{katz_sphinx_2023}, both of which can be utilised to improve the fidelity of the forward modelled emission.

\begin{figure}
    \includegraphics[width=\columnwidth]{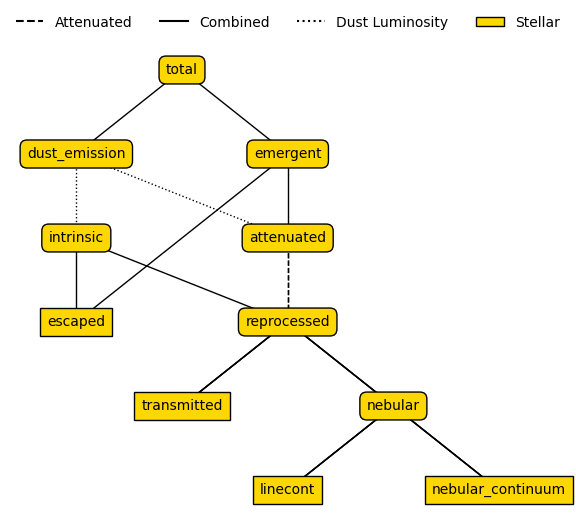}
    \caption{A tree diagram of the Pacman emission model. Each named item represents an SED object that can be extracted from the emission model. Square boxes indicate extractions, rounded boxes indicate the results of transformations, combinations or generations. Solid lines indicate combination relations between outputs, dashed lines represent attenuation transformations, and dotted lines generation dependencies.}
    \label{fig:pacman_tree}
\end{figure}
 
In order to permit flexibility in these various approaches, we template out the emission pipeline.
We then employ a dependency injection pattern when generating emissions, where the template is a dependency defining what will be generated when it is passed to a galaxy or component, and what the stages in that generation will be.
These templates are called \emodels. In their simplest form, an \emodel defines the mapping of a set of inputs to a single output emission by following a simple set of rules, e.g. feeding in an SPS model \grid and a stellar component to get the incident emission from the population of stars, or inputting an intrinsic emission and a dust curve to get the dust attenuated emission. By chaining together these individual steps, \emodels can become arbitrarily complex, leading to a network of connected \emodels, each producing a single emission that leads into the next step. For an example of one of these networks (or trees), see \fig{pacman_tree}. 

\begin{figure*}
	\includegraphics[width=\textwidth]{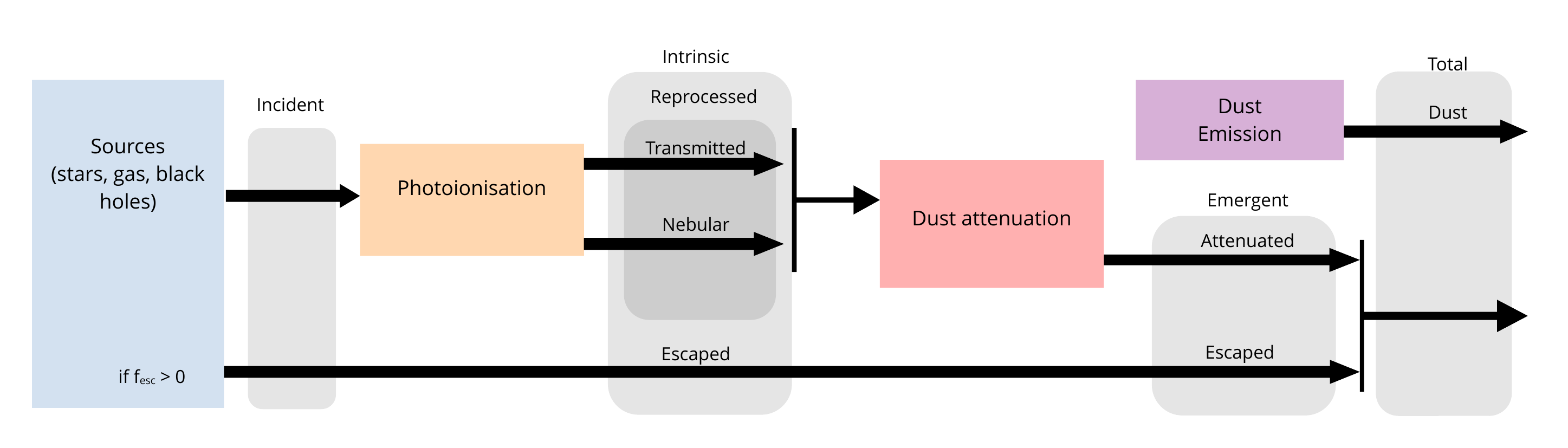}
    \caption{A flowchart showing the typical procedure used to generate the composite emission from a galaxy, and the standardised names for various outputs at each stage. These include the \textit{incident} radiation from sources of emission, that feeds into photoionisation codes to produce the \textit{intrinsic} emission (and various sub-divisions therein), to the \textit{emergent} emission after dust attenuation, and \textit{total} including dust emission.}
    \label{fig:flowchart}
\end{figure*}

Once an \texttt{EmissionModel} is constructed it can be used to generate emission from a galaxy or a component.
\texttt{EmissionModel}s are agnostic to whether they are producing spectra, lines or even images; the details of each of these cases is handled outside an \texttt{EmissionModel}.
They are also agnostic to whether they are producing spectra for particle or parametric galaxies or components.
Once an \texttt{EmissionModel} is initialised any parameters can be subsequently modified, new operations added or relabeled, or existing operations replaced with one or more new operations.
The various spectra produced during the processing of an \texttt{EmissionModel} can optionally be saved, allowing the user to inspect intermediate steps, or the emission from specific physical properties.
Finally, \texttt{EmissionModel}s can be combined; a key example of this is combining a stellar emission model and an AGN emission model to produce the composite spectra for an entire galaxy.
Below we describe the main operations in an emission model in detail (\sec{em_operations}). 
We also describe some fiducial emission models provided with \textsc{Synthesizer} in \app{fiducial_emodel}, as well as our standard naming convention in \app{em_naming}.
A flowchart showing a typical work flow, with the standard naming convention, is shown in \fig{flowchart}.

\vspace{1.0cm}

\subsection{Operations}
\label{sec:em_operations}

An \emodel can assume one of four simple `operations': \texttt{Extraction}, \texttt{Generation}, \texttt{Transformation}, and \texttt{Combination}.
An \emodel can only perform one of these operations, and each produces a singular output. 
These operations are described in detail below.

Any operation can also be combined with a mask; masks can take two forms.
The first masks the wavelengths of the spectra the \emodel can produce; this is typically only done to improve performance, since this reduces the number of operations carried out throughout an \emodel.
The second type of mask is based on an attribute on a component, that can be used to to isolate a specific subset of the emitter.
For instance, a mask could be defined for stellar age, to produce emission only for young stellar populations, or based on 2D projected radii to get emission within an aperture.
These property masks can use any attribute of the component, and any number of these can be applied within a single \emodel.
Note that, regardless of these masks, every emission produced by a root model will be the same shape.

\subsubsection{\texttt{Extraction}}
An \texttt{Extraction} defines the `extraction' of an emission from a \grid. More specifically, an \texttt{Extraction} defines the process of mapping component properties onto a \grid of emissions. It thus requires a component (i.e. any \texttt{Stars}/\texttt{BlackHoles}/\texttt{Gas} component) and a suitable \grid as input, along with any required masks.

The axes of the \grid (and thus the component properties mapped onto it) can be entirely arbitrary, but every grid axis must have a corresponding component attribute to extract.
As discussed in \sec{grids}, for SPS models, these axes are typically stellar age and birth metallicity, but there is no limit to the number of axes, nor what those axes are.
The mapping onto the \grid can be done using either a cloud-in-cell (CIC) or nearest grid point (NGP) approach (these methods are further detailed in \sec{part-spec-gen}).

\texttt{Extractions} in \fig{pacman_tree} are denoted by square nodes at the leaves of the tree. \texttt{Extractions} are always leaves since they are the only operation that never requires another model as input.

\subsubsection{\texttt{Generation}}

Sometimes the emission can be predicted (or at least approximated) using simple analytic prescriptions, without having to resort to grids.
In these cases we treat the model as a \textit{generator}, to distinguish it from the more comprehensive and flexible \texttt{grids}.
The emission from a generator can still be linked to the properties of a component.


The thermal emission from dust can be approximated by a simple modified black body.
\textsc{Synthesizer} provides access to these analytic forms, with a range of sophistication in how they are linked to the physical properties of galaxies.
More details on the provided analytic forms of the dust emission are provided in \sec{dust_emission}.
Dust emission \texttt{Generations} derive information from existing models, placing them away from the leaves in the emission tree hierarchy.
This can be seen in \fig{pacman_tree} where a \texttt{dust\_emission} model is linked by dotted lines to its intrinsic and attenuated models, which are required to apply the energy balance technique and derive the dust luminosity.


A \texttt{Template} is not shown in \fig{pacman_tree}, but like \texttt{Extractions} they do not require another model as an input.
A \texttt{Template} \texttt{Generation} would therefore appear as a round-cornered leaf in an emission tree.
Example of template generators could include AGN templates.

\subsubsection{\texttt{Transformation}}

A transformation is an emission model operation that, at its most general, applies a linear transformation to the output of an emission model.
More specifically, a transformation could represent the wavelength-dependent attenuation of a spectrum, for example, or the resonant boosting of a particular line (e.g. Lyman-$\alpha$).
Transformation operations are indicated with dashed lines in an emission tree, and create round-cornered elements in an emission tree.

For attenuation operations you need:
\begin{itemize}
    \item The dust curve to apply (\texttt{dust\_curve}, see \sec{dust_curves}).
    \item The emission model output to apply the attenuation to (\texttt{apply\_dust\_to}).
    \item The assumed V-band optical depth (\texttt{tau\_v}).
\end{itemize}

\subsubsection{\texttt{Combination}}
A \texttt{Combination} operation is the simplest possible operation. It defines the `combination' (or addition) of two or more emissions. It only requires the models whose emissions will be combined as input. \texttt{Combinations} can be seen in \fig{pacman_tree} anywhere two models are connected by solid lines to their parent model.

\section{Generating Spectra}
\label{sec:generating_spectra}

\begin{figure*}
	\includegraphics[width=\textwidth]{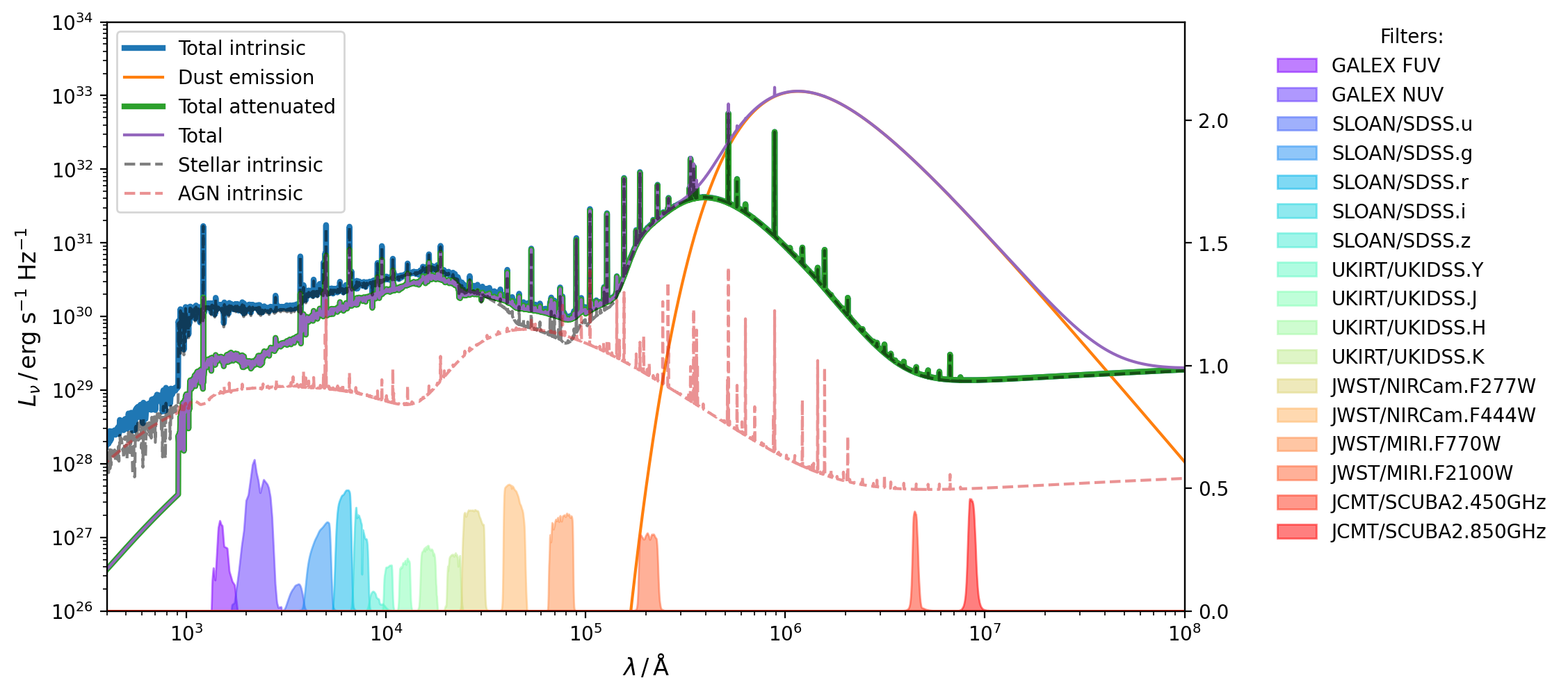}
    \caption{Composite galaxy spectra example.
    The stellar component has a mass of $10^{11} \, {\rm M_{\odot}}$,  and assumes a log-normal star formation history ($\tau = 1$, peak age = 400 Myr, max age = 2 Gyr) and solar metallicity ($Z = 0.014$).
    The emission is obtained using the Pacman model, assuming an escape fraction, $f_{\rm esc} = 0.1$, using the BPASS v2.3 SPS model with the fiducial photoionisation parameters.
    The AGN component assumes a NLR / BLR covering fraction of 0.1, ionisation parameter of 0.1, and a Hydrogen density of $10^5$ cm$^{-3}$.
    Both components are attenuated by a diffuse ISM component assuming a MWN18 attenuation curve, with $\tau_{\rm V} = 0.33$.
    The young ($\leqslant 10 \, {\rm Myr}$) stellar component is further attenuated by a factor of 0.67.
    Dust emission is modelled using a Greybody, with temperature 30 K and emissivity of 1.2.
    }
    \label{fig:galaxy_spectra}
\end{figure*}

Once a user has chosen a grid, created a galaxy object (consisting of various components), and defined an emission model, synthesizer can be used to generate spectral energy distributions, or spectra.
Spectra are the primary base observable from which most other observables are derived (see \sec{emissions}).

An example composite SED, with each components contribution to the overall emission, is shown in \fig{galaxy_spectra}.
Below we describe how synthesizer calculates the SED from particle and parametric galaxy objects.
Semi analytic models represent a special case, and are described in \app{sam_spectra}.

\subsection{Particle Based Models}
\label{sec:part-spec-gen}

Particle based models, such as hydrodynamic simulations, model stars and black holes as discrete particles that represent a single stellar population, or a single black hole.
These individual particles have a number of parameters that can control the emission.
For stars, the primary parameters are the age and metallicity, which can be associated with the emission from a grid object.
In order to predict the composite emission from a collection of particles, synthesizer uses a grid assignment scheme to find the overall contribution of each star particle to the grid \textit{weights}.


Synthesizer implements two look up schemes.
The Nearest-Grid-Point (NGP) scheme simply finds the closest value in the grid space, and assigns the spectra from this point to that particle.
The Cloud-In-Cell (CIC) scheme combines the spectra from the nearest points in the grid space, weighted by their proximity.
This approach assumes a smooth, linear evolution between grid points, which may not be the case for certain spectral features; users should take care that the resolution of their chosen grid suits the fidelity of their application.

In both schemes, the spectra assigned to a given particle depends on whether the grid dimensions are described linearly or logarithmically.
Users can specify this choice for each dimension as required, however recommended defaults are saved in each grid file.
For age and metallicity we recommend using logarithmic scaling.
For all grids, we truncate the predictions to the edge of the grid, and do not extrapolate beyond the grid boundaries; users should take care that either their input stellar properties are covered by the chosen grid, or that they trust the predictions at the grid boundaries for out of distribution particles. This is also the case for parametric models, described below.

\subsection{Parametric Models}

Parametric models implement a slightly different look up scheme when combined with grids.
This is to account for the fact that, rather than discrete Simple Stellar Populations, parametric models represent the star formation history in bins, or smooth parametric forms.
For parametric models, synthesizer integrates the star formation history (SFH) over the grid dimensions to determine the contribution of each grid point.

\vspace{1.0cm}

\section{Observables and Instruments}
\label{sec:emissions}

\subsection{Spectral Energy Distribution Objects}
\label{sec:sed}

A Spectral Energy Distribution, or SED, describes the energy distribution of an emitting body as a function of frequency / wavelength.
There are a number of different ways to generate SED's in \textsc{Synthesizer}, but in every case the resulting SED is always stored in an \texttt{Sed} object.
\texttt{Sed} objects can be combined or modified using the overloaded addition and division operators.
They can also be resampled onto new wavelength grids, using the \textsc{Spectres} package \citep{carnall_spectres:_2017}.
\texttt{Sed} objects can contain multiple spectra (e.g. for multiple galaxies or particles), and operations will be applied utilising vectorisation.
Figures \ref{fig:mosaic} and \ref{fig:galaxy_spectra} show spectra for two example galaxies, split into multiple different sources and components.

The Sed class contains a number of methods for calculating derived properties.
The bolometric luminosity can be computed using a trapezoidal numerical integration method by default, but alternative integration routines can be specified.
The luminosity in a given window can be calculated, given arbitrary wavelength or frequency units, as well as the strength of a spectral break if given two windows, or the slope.
We provide specific methods for calculating the Balmer and D4000\textup{~\AA} breaks, and the UV $\beta$-slope.
The ionising photon production rate can also be calculated, given an ionisation energy (defaults to atomic Hydrogen at 13.6 eV).

\texttt{Sed} objects are defined by default in the rest frame, however the observed flux at different redshift's can be calculated given an astropy cosmology object \citep{robitaille_astropy:_2013} and a redshift.
The attenuation due to dust or the IGM can also be calculated; further details are provided in \sec{dust} and \app{igm_absorption}, respectively.
Photometry can be calculated directly from Sed objects, and is described in \sec{photometry}.

\subsection{Emission Line Collection Objects}
\label{sec:lines}

Similar in construction to the \texttt{Sed} object, a \texttt{LineCollection} object describes the properties of a collection of emission lines.
Like spectral energy distributions, line collections can be extracted directly from \texttt{Grid} objects, or generated from a \texttt{Galaxy} or its components (i.e. Stars, Gas, or BlackHoles).

Line IDs follow the same convention as in \textsc{Cloudy}, whereby a line is usually represented as
\begin{align*}
    &{\rm \{atomic/molecular \; notation\}\{ionisation \; state\}} \\
    &{\rm \{wavelength\}}
\end{align*}
The ionisation state is in the usual astronomical notation, e.g. H 1 for atomic hydrogen, but different for molecules, with the state denoted by `+’ or `-’ (e.g. HCO+).
Individual lines can be accessed much like a dictionary, as well as composite lines by passing a string of comma separated IDs. 
We also provide aliases to a number of common individual and composite lines.

We provide methods for computing line ratios, plotting common diagnostic diagrams (e.g. BPT[-NII]), blending lines (given some assumed spectral resolution), and for combining and scaling line collections.

\vspace{1.0cm}

\subsection{Filters}

Photometric filters in synthesizer are defined using two dedicated objects.
\texttt{Filter} objects describe individual filters, whereas \texttt{FilterCollection} objects describe collections of Filters that behave like a list and a dictionary, with extra attributes and methods to efficiently work with multiple Filter objects.
Filters can be used for producing photometry from Sed objects, as well as for creating monochromatic or RGB images.

\textsc{Synthesizer} provides a number of different ways to define a Filter or set of Filters.
\textit{Generic} filters can be defined with user provided arrays for the wavelength and transmission.
\textit{Top Hat} filters can be defined given a minimum and maximum wavelength, or an effective wavelength and Full Width Half Maximum (FWHM); the transmission is set to one in this range and zero outside.
Finally, \textsc{Synthesizer} provides access to the Spanish Virtual Observatory (SVO) filter service\footnote{\url{http://svo2.cab.inta-csic.es/theory/fps/}}, which provides filter transmission curves for a number of observatories and instruments \citep{rodrigo_svo_2012,rodrigo_svo_2020,rodrigo_photometric_2024}.
This requires an internet connection.
The user must provide the filter code in “Observatory/Instrument.code” format (as shown on the SVO website).
A number of transmission curves obtained from SVO for a range of different instruments are shown in \fig{galaxy_spectra}.

\texttt{FilterCollections} can be saved to a HDF5 file and loaded at a later date, which can be useful for using SVO filters on machines where an internet connection is not available, such as in cluster environments.

\subsection{Photometry}
\label{sec:photometry}

Once spectra have been computed (from a galaxy or component, or extracted from a grid) we can produce rest-frame and observer-frame photometry by combining the \texttt{Sed} object containing the (multidimensional) spectra with a \texttt{FilterCollection} defining the transmission curve of a set of filters.
Doing so produces a \texttt{PhotometryCollection} containing the luminosities / fluxes in each band (with units), and methods for manipulating and visualising them.
Photometry can also be produced at the particle level, allowing for the creation of images and data cubes (see \sec{imaging}).
Given photometry at the particle level, \textsc{Synthesizer} also provides methods for calculating fractional light radii.

\textsc{Synthesizer} assumes the AB magnitude system and a photon count detector\footnote{Other photometric systems will be introduced in a future version, as well as support for energy detectors.} \citep[see][]{fouesneau_pyphot_2025}.
If we consider a filter transmission profile $T(\lambda)$, the number of photons arriving on the filter is given by
\begin{align}
    N \,/\, {\rm s^{-1} \, cm^{-2}} = \frac{1}{hc} \int_{\lambda} f_{\lambda} \lambda T(\lambda) \, d\lambda
\end{align}
where $f_{\lambda}$ is the wavelength dependent flux density incident on the filter, in units of ${\rm erg / s / cm^2 / \textup{~\AA}
}$.
For frequency dependent spectral flux density we convert $f_{\nu} = \lambda f_{\lambda} / c^2$.
The mean of the flux density is then given by
\begin{align}
    \overline{f_{\lambda}}(T) \,/\, ({\rm erg \, s^{-1} \, cm^{-2} \, \textup{~\AA}
^{-1}}) = \frac{\int_{\lambda} \lambda f_{\lambda} T(\lambda) d\lambda}{\int_{\lambda} \lambda T(\lambda) d\lambda}
\end{align}
Where we are calculating the luminosity in a given band, we assume the standard reference distance of 10 parsecs, and convert the flux density units appropriately.

If we define the pivot wavelength $\lambda_p$ as $\overline{f_{\nu}} = (\lambda_p^2 \,/\, c) \overline{f_{\lambda}}$, then the AB magnitude is given by
\begin{align}
    m_{\rm AB} (T) = -2.5 {\rm log_{10}} \, \overline{f_{\nu}} - 48.60 \;\;.
\end{align}

\subsection{Imaging and Data Cubes}
\label{sec:imaging}

\textsc{Synthesizer} can be used to generate spectral data cubes and photometric images.
These can be generated from particle and parametric data.
The functionality is contained in the following objects:
\begin{itemize}
    \item \texttt{SpectralCube}s, containers for visual spectral (spaxel) data cubes.
    \item \texttt{Image}s, for individual images.
    \item \texttt{ImageCollection}s, for multiple \texttt{Images}.
\end{itemize}
\textsc{Synthesizer} can also be used to create maps of specific physical properties, described in \app{property_maps}.

When creating an \texttt{Image}, the user must provide a resolution and Field of View (FoV), where the resolution defines the number of subdivisions for each pixel along each spatial dimension, and the FoV is in physical spatial units in the source frame.
Images can be made that include multiple components (e.g. stars + black holes).
In practice, \textsc{Synthesizer} creates two separate images for each component, and then combines them to make a composite image.

When creating an \texttt{Image} derived from particle data, the spatial extent of each particle can be controlled.
Either each particle is treated as a point source, which are then summed over the image through a simple histogram, or each particle is distributed over a 2D kernel in the image plane.
In the latter a kernel must be assumed; a number of 2D and 3D kernels are provided (see \app{kernels}).

\subsubsection{RGB Images}
RGB images can be quickly constructed by supplying three different bands corresponding to the red, green and blue channels, which will be combined to produce the full colour RGB image.
An example of this is shown in \fig{mosaic}.
Alternatively, multiple bands or maps (e.g. from emission lines) can be combined within each colour channel, as long as the data is constructed as an HxWx3 array, where H is the image height in pixels, W is the image width in pixels, and the 3 colour channels, respectively.
Synthesizer currently only supports equal weighting of each RGB channel; alternative schemes, such as the \cite{lupton_preparing_2004} scheme, will be supported in a future release.

\subsubsection{Point Spread Function (PSF)}

To properly model observations from a particular instrument we must take into account the point spread function (PSF).
PSFs can be sourced however the user wishes\footnote{For example, we recommend the \textsc{webbpsf} package (see \href{here}{https://www.stsci.edu/jwst/science-planning/proposal-planning-toolbox/psf-simulation-tool}) for the James Webb Space Telescope \citep{perrin_updated_2014}.}; \textsc{Synthesizer} accepts simple arrays describing the form of the PSF.
To get accurate results from PSF convolution it is recommended to use a super-sampled image (i.e. much higher resolution than the PSF).

\fig{mosaic} shows an example RGB image including the effect of the JWST PSF.

\subsubsection{Noise Modelling}

The final ingredient for a fully forward modelled synthetic image is a noise model.
\textsc{Synthesizer} provides 4 different approaches for modelling noise:
\begin{itemize}
    \item Apply a noise array: Add an existing noise field / array.
    \item Apply noise from std: Derive a noise distribution, centred on 0, given a user specified standard deviation, and then generate and add a noise array.
    \item Apply noise from snr (aperture): Derive a noise distribution from a Signal-to-Noise Ratio (SNR), defined in an aperture with size aperture radius and a specified depth. This will derive the standard deviation of the noise distribution assuming for an aperture, before deriving the per pixel noise, computing the noise array and adding it.
    \item Apply noise from snr (point source): Derive a noise distribution from a SNR and depth. This will derive the standard deviation of the noise distribution assuming $SNR = S / \sigma$ for a pixel before computing the noise array and adding it. This behaviour can be achieved by omitting aperture radius in the call to apply noise from snr.
\end{itemize}
As with applying a PSF, these methods have singular versions (as listed above) which can be used on an individual \texttt{Image}, and pluralised versions which can be used on an \texttt{ImageCollection}, and take dictionaries for each of the arguments.
If an \texttt{Image} has units then the passed noise array or noise standard deviation must also have units.

Applying noise with any of the methods described above will return a new ImageCollection / Image containing the noisy image, and the noise array and weight map stored in attributes on the object.
\fig{mosaic} shows an example RGB image including the effect of a simple noise model.

\subsubsection{Spectral Data Cubes}

Spectral data cubes are representations of HxWxL arrays, where H and W are the pixel height and width, and L is the number of wavelength points.
\textsc{Synthesizer} allows users to create these high dimensional data products, using either a histogram or smoothed approach as for 2D imaging.
The procedure is much the same as for imaging, and can be defined either using observer frame fluxes or luminosities, but the user must provide the wavelength grid over which the data cube will be constructed. 

\textsc{Synthesizer} provides helper methods for constructing animations of data cubes that iteratively loop through the wavelength axis; examples of these can be found in the documentation.

\section{Pipeline}
\label{sec:pipeline}

The \texttt{Pipeline} class is a high-level interface that allows users to easily generate observations for a given catalogue, emission model, and set of instruments.
Users simply set up the \texttt{Pipeline} object, attach the galaxies, and run the observable methods desired. These include, spectra, emission lines, photometry, images and data cubes.
The \texttt{Pipeline} will generate all the requested observations for all (compatible) instruments and galaxies, before writing them out to a standardised HDF5 format.
Abstraction into the \texttt{Pipeline} class allows for hybrid parallelization over local threads as well as MPI.
For now, load data partitioning is left to the user, but once the data is loaded all operations can be hybrid parallelised.
If parallel h5py has been installed then outputs can be collectively written to a single hdf5 file; if not, individual files are output and can be combined on exit.  
\section{Parallelism \& Scaling}
\label{sec:parallelism_scaling}

\textsc{Synthesizer} aims to be flexible in its approach to parallelism, enabling the user to use whatever paradigm fits their use case (e.g. shared memory, distributed or hybrid). 
We choose to leave distributed memory parallelism (MPI), in the hands of the user\footnote{With the exception of the Pipeline class, which can take an MPI communicator object as an argument and then coordinate a distributed calculation internally (See \sec{pipeline} for more details).}, while shared memory parallelism can be optionally invoked at a function call level for specific, intensive computations.
This allows the user to have fine-grained control over parallelism in their code, and to avoid overheads where parallelism is not required.

\begin{figure}
	\includegraphics[width=\columnwidth]{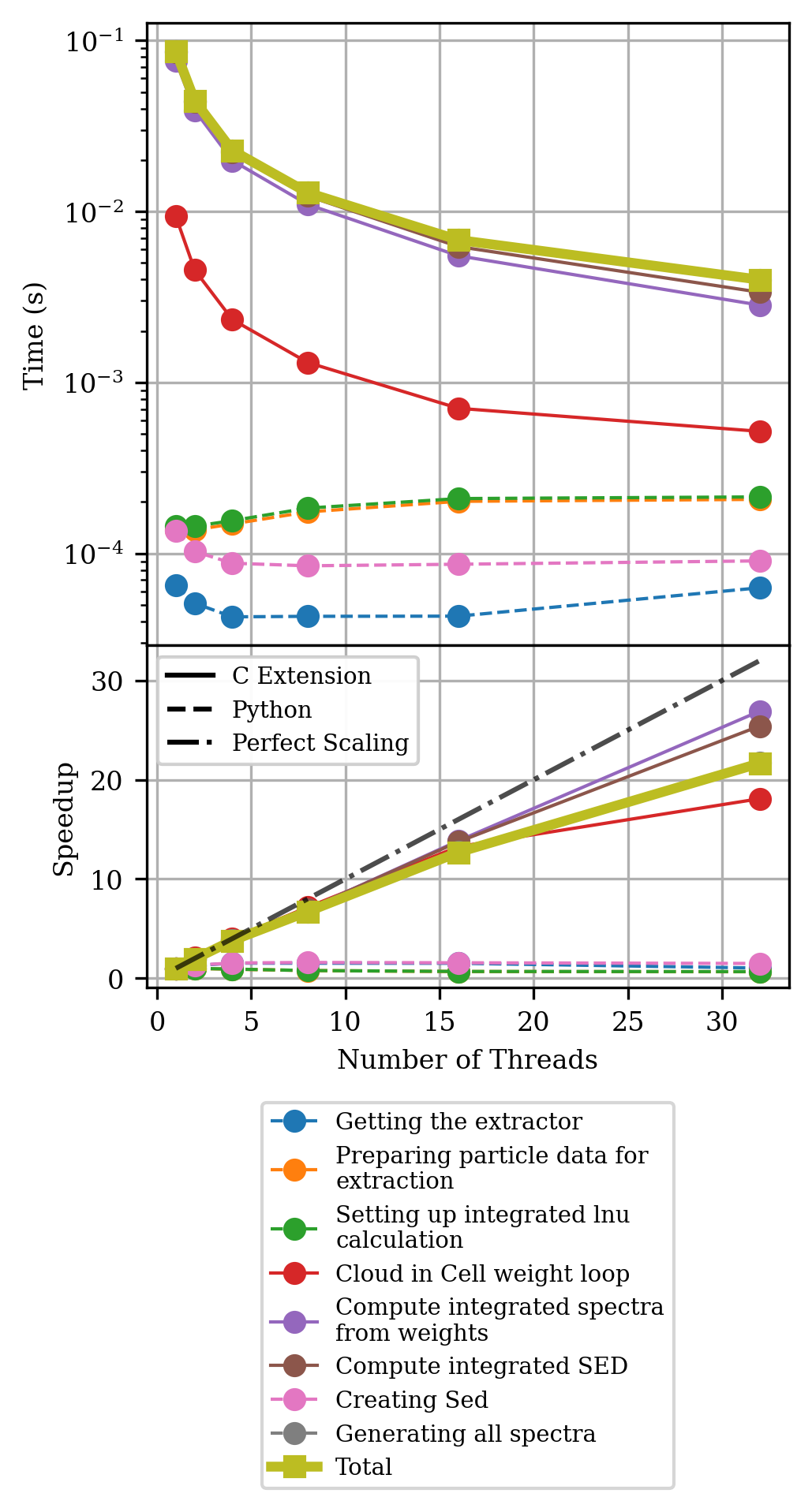}
    \caption{The shared memory scaling of integrated spectra generation with the number of threads used (for a simple single spectra model). The stellar component used to generate this spectrum had $10^5$ stellar particles and generated spectra with 9244 wavelength elements (though this scaling is replicated for a range of particle counts and wavelength resolutions). This test was run on a High Performance Computing centre compute node with 2 AMD Epyc Milan CPUs, each with 64 physical cores, 4 NUMA regions, and 1TB of RAM (though only a fraction of this memory was used). Only operations that took more than 1\% of the total time taken by the maximum number of threads are plotted. The lower legend denotes different operations performed during the calculation.
    Solid lines denote operations in C, dashed lines denote Python operations, and the dotted and dashed line denotes perfect scaling.}
    \label{fig:integrated_spectra_scaling}
\end{figure}

\begin{figure}
	\includegraphics[width=\columnwidth]{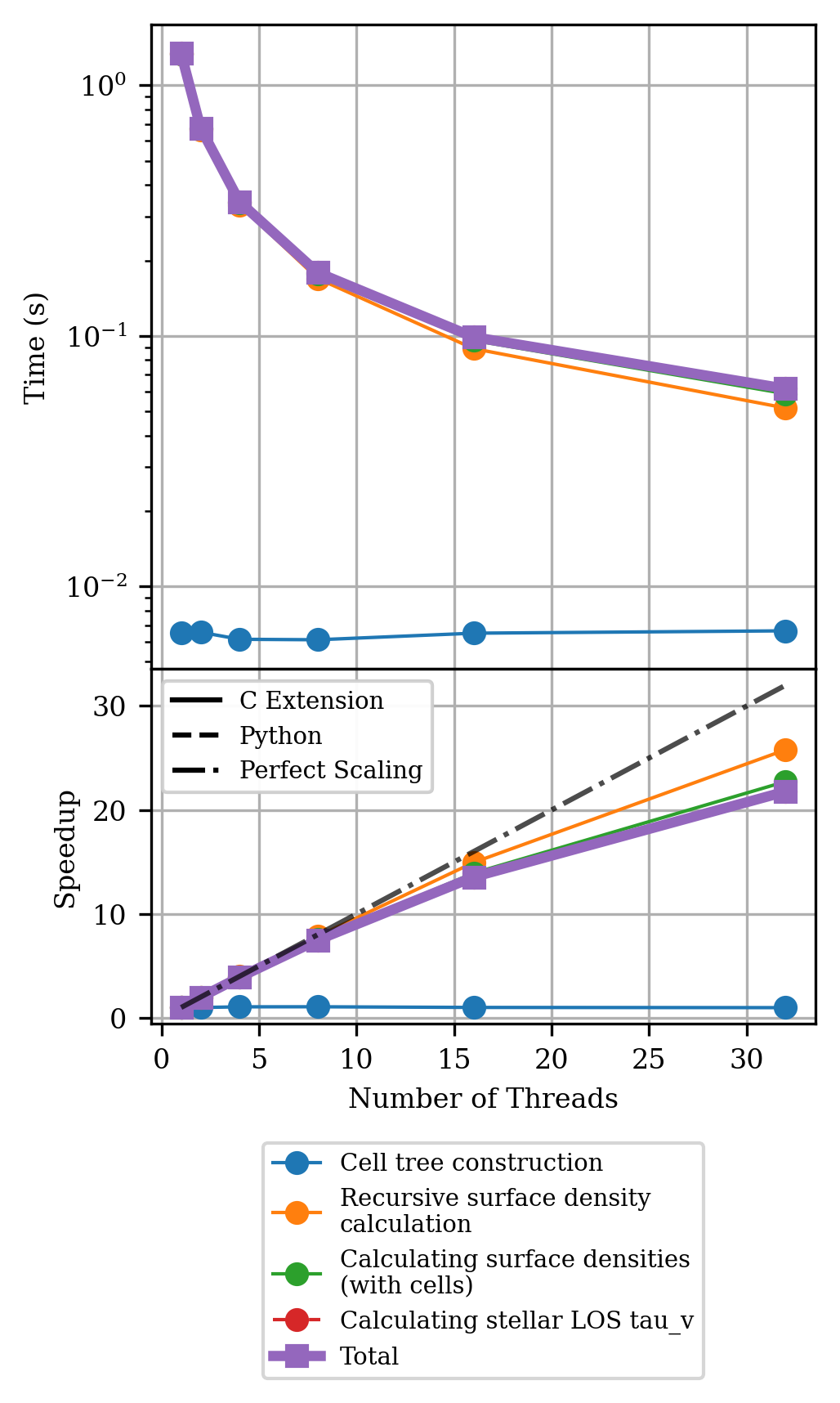}
    \caption{The scaling of the line-of-sight column density calculations with the number of threads used. This calculation was performed with $10^5$ gas particles and $10^5$ stellar particles (though the scaling is the same across a wide range of particle counts). The test was run on a High Performance Computing centre compute node with 2 AMD Epyc Milan CPUs, each with 64 physical cores, 4 NUMA regions, and 1TB of RAM (though only a fraction of this memory was used). Only operations that took more than 1\% of the total time taken by the maximum number of threads are plotted. The lower legend denotes different operations performed during the calculation.
    Solid lines denote operations in C, dashed lines denote Python operations, and the dotted and dashed line denotes perfect scaling.}
    \label{fig:los_scaling}
\end{figure}

To avoid the drawbacks of the Python Global Interpreter Lock (GIL) for threaded tasks, we use OpenMP in the C extensions used for spectra generation, integration, Line-Of-Sight (LOS) surface density calculations, imaging, and other computationally intensive tasks.
Making use of these threadpools is as simple as passing the \texttt{nthreads} argument to the relevant function.
The performance of the code will scale with the number of threads used, up to the number of physical cores on your machine.
Scaling performance test results for integrated spectra generation are shown in \fig{integrated_spectra_scaling} and line of sight column density calculation results are shown in
\fig{los_scaling}. In both cases, the calculations scale extremely well for moderate threading number increases. However, there's a noticeable decline in performance above 16 threads evident in both plots. This is expected due to the increasing overhead in global reductions, but is also exacerbated by the fact that each NUMA region on the hardware used for testing has 16 threads. Exceeding 16 threads on this hardware incurs extra overheads as the boundary between NUMA regions is crossed.

\section{Summary \& Conclusions}
\label{sec:conc}

We have introduced \textsc{Synthesizer}\footnote{\url{https://synthesizer-project.github.io}}, a fast, modular, flexible and extensible code for creating synthetic astronomical observables.
Synthesizer is free and open source software (FOSS), with complete documentation, and a comprehensive unit and integration test suite.
We have introduced the historical background and motivation for a code like \textsc{Synthesizer}, to enable both forward and inverse modelling approaches using a consistent pipeline, as well as a range of modelling choices not typically available in other similar codes.
Synthesizer is not intended to replace high fidelity approaches, but instead as a platform for generating rapid predictions for different modelling assumptions, in order to assess their impact on observables, and for generating large training sets for modern deep learning techniques.

We have described the main components of \textsc{Synthesizer}, including:
\begin{itemize}
    \item \textsc{Galaxy} objects, and their constituent stellar, gas and black hole components, and how these can be defined in terms of particle and parametric forms.
    \item \textsc{Grids}, of both stellar population synthesis and active galactic nuclei emission, as well as photoionisation processing.
    \item \textsc{EmissionModels}, a synthesizer modular concept allowing the constructing of complex models of various components, and how they interact when predicting the multi-wavelength emission from a galaxy.
    \item We described how grids, galaxies (and / or their components) and emission models can be combined to generate spectra.
    \item We describe a number of observables that can be generated with \textsc{synthesizer}, including high resolution spectra, emission line collections, photometry, imaging and data cubes.
    \item We introduce the \textsc{Pipeline}, a high level interface for combining various synthesizer functionality in just a few lines of Python code.
    \item We describe the scaling performance of the code, and various aspects that have been natively parallelised.
\end{itemize}

A code like \textsc{Synthesizer} has many evident use cases.
Users that wish to forward model key observable distribution functions from cosmological simulations, such as luminosity functions and colours, can now do so for different SPS, dust and photoionisation models, simply by changing a few small parameters in the pipeline.
This unlocks the possibility of robustly assessing the dependence of the predictions on the forward model, providing quantitative baseline uncertainties on these quantities, that can be compared to the uncertainties from inverse modelling approaches.
The inclusion of AGN and stellar emission in a single unified code also opens up the possibility to assess the relative contribution of these components to galaxy emission at different wavelengths and across cosmic time.
Finally, the speed of \textsc{Synthesizer} opens up a number of opportunities in Simulation Based Inference (SBI) schemes; an example of this has already been demonstrated in \citep{lovell_learning_2024}, by applying to the large suite of CAMELS simulations. 
Numerous other applications of \textsc{Synthesizer} exist that we cannot address here due to space limitations; we look forward to seeing the community explore both these applications and creative uses the authors have not yet envisaged.
    
There are a number of avenues for future development of \textsc{Synthesizer}.
High dimensional grids can consume large amounts of disk space; to overcome this, emulators can be used to model the emission as a function of grid parameters, reducing the disk requirements and allowing for rapid high dimensional grid exploration \citep[see][]{alsing_speculator:_2020,lovell_sengi_2021}.
Emulators can also be used to predict the results of dust RT simulations, to allow for rapid predictions informed by the highest fidelity models \citep[e.g.][]{sethuram_emulating_2023}, complementing the relatively simpler dust models provided to date.

Auto differentiability is rapidly being implemented in a number of astrophysical codes to allow for gradient based inference and optimisation schemes (e.g. Pandya et al. \textit{in prep.}).
Whilst the class based structure of \textsc{Synthesizer} does not lend itself naturally to such approaches, we hope to explore the implementation of autodiff in various parts of the code to allow for these schemes.

We actively encourage the community to use and develop the code\footnote{Documentation is available at \url{https://synthesizer-project.github.io}, and the repository at \url{https://github.com/synthesizer-project/synthesizer}}, and are very happy to receive issues and pull requests for new functionality.

\section*{Acknowledgments}
The authors wish to thank Hollis Akins, Sabrina Berger, Connor Sant Fournier, Thomas Harvey, Kartheik Iyer, Marco Leonardi, Borja Pautasso, Ashley Perry, Paurush Punyasheel, and Laura Sommovigo for contributions to the code base.
We thank Tjitske Starkenburg and Thomas R. Greve for useful discussions, the Illustris-TNG and CAMELS teams for making their simulation data public, and Rachel Somerville and Aaron Yung for providing the SC-SAM data.
Finally, we would like to thank Peter Thomas for his warm words, encouragement and motivation -- it's finally out Peter, and only a year late!

This work used the DiRAC Memory Intensive service (Cosma7 \& Cosma8) at Durham University, managed by the Institute for Computational Cosmology on behalf of the STFC DiRAC HPC Facility (www.dirac.ac.uk). The DiRAC service at Durham was funded by BEIS, UKRI and STFC capital funding, Durham University and STFC operations grants. DiRAC is part of the UKRI Digital Research Infrastructure.
This research has made use of the SVO Filter Profile Service ``Carlos Rodrigo", funded by MCIN/AEI/10.13039/501100011033/ through grant PID2023-146210NB-I00.
This work was supported by the Simons Collaboration on “Learning the Universe”.
WJR, APV and SMW acknowledge support from the Sussex Astronomy Centre STFC Consolidated Grant (ST/X001040/1).
APV acknowledges support from the Carlsberg Foundation (grant no CF20-0534). 

We list here the roles and contributions of the authors according to the Contributor Roles Taxonomy (CRediT)\footnote{\url{https://credit.niso.org/}}.
\textbf{Christopher C. Lovell, William J. Roper, Aswin P. Vijayan, Stephen M. Wilkins}: Conceptualization, Data curation, Formal Analysis, Methodology, Project administration, Resources, Software, Validation, Visualization, Writing - original draft, Writing - review \& editing,
\textbf{Sophie Newman, Louise Seeyave}: Data curation, Validation, Software, Visualization, Writing - review \& editing.

\bibliographystyle{mnras}

\bibliography{synthesizer,sophie,non_zotero,extra}

\begin{appendix}

\section{Units}\label{sec:units}

To ensure unit consistency and capture unit-related errors early, we implement a package-wide unit system underpinned by the \unyt \footnote{\url{https://unyt.readthedocs.io/en/stable/}} Python package \citep{goldbaum_unyt_2018}. The machinery governing this unit system is housed in the \texttt{units} module.  
The units module contains the \texttt{Units} class, which defines the unit system itself, and the \texttt{Quantity} descriptor class. 

The \texttt{Units} object holds the units of all unit-carrying attributes in \synthesizer. This object utilises a \texttt{Singleton} design pattern, which means only a single instance of `Units` can exist within a single Python interpreter instance\footnote{A Python interpreter instance is essentially a single running copy of the Python runtime environment. It is created each time you run Python (for example, when you launch the Python shell or execute a script). All code executed within that session shares the same runtime.}. We utilise this pattern to ensure there can only ever be a single unit system. 

The \texttt{Units} object itself doesn't need to be directly instantiated by the user, it is created in the background the first time a \texttt{Quantity} is interacted with. Once instantiated, the unit system is fixed. This protects against any potential confusion introduced if units are changed part way through an analysis. However, the user is free to modify the unit system before the first \texttt{Units} instance is created. This can be done by explicitly instantiating a \texttt{Units} instance with a dictionary of modified units; as before the unit system is then fixed from this instantiation onwards.

The default unit system itself is stored in a \texttt{yaml} file inside the package. This default unit system defines the units for different `categories' of attributes which are then loaded and attached during \texttt{Units} instantiation. The user can permanently modify the units defined in this file (or reset them to the original system) either by modifying the file \textit{in situ}, or by using methods on the \texttt{Units} instance to save its current state. This enables the user to tailor the defaults to their science case without having to modify the unit system every time they use it. The default unit system is detailed in Table \ref{tab:unit-system}.

\begin{table*}
  \centering
  \caption{The default units system}
  \label{tab:unit-system}
  \begin{tabular}{l|c|l}
    \hline
    \textbf{Category} & \textbf{Unit} & \textbf{Example Attributes} \\
    \hline
    \texttt{spatial} & Mpc & \texttt{Stars.coordinates} ($\bar{x}$), \texttt{Gas.smoothing\_lengths} ($h$) \\
    \texttt{mass} & $M_\odot$ & \texttt{Stars.initial\_masses} ($M_{\rm ini}$), \texttt{Stars.current\_masses} ($M_\star$), \texttt{BlackHole.mass} ($M_\bullet$) \\
    \texttt{cosmic\_time} & yr & \texttt{Stars.ages} ($T_{\rm age}$) \\
    \texttt{velocity} & km/s & \texttt{Gas.velocities} ($\bar{v}$) \\
    \texttt{temperature} & K & \texttt{Gas.temperatures} ($T$) \\
    \texttt{angle} & degree & \texttt{BlackHoles.inclinations} ($i$), \texttt{Sersic2D.theta} ($\theta$) \\
    \texttt{wavelength} & \AA & \texttt{Sed.lam} ($\lambda$) \\
    \texttt{frequency} & Hz & \texttt{Sed.nu} ($\nu$) \\
    \texttt{luminosity} & erg/s & \texttt{LineCollection.luminosities} ($L$), \texttt{BlackHoles.bolometric\_luminosity} ($L_{\rm bol}$) \\
    \texttt{luminosity\_density\_frequency} & erg/Hz/s & \texttt{Sed.lnu} ($L_\nu$) \\
    \texttt{luminosity\_density\_wavelength} & erg/s/\AA & \texttt{Sed.llam} ($L_\lambda$) \\
    \texttt{flux} & erg/cm$^2$/s & \texttt{Line.flux} ($F$) \\
    \texttt{flux\_density\_frequency} & nJy & \texttt{Sed.fnu} ($F_\nu$) \\
    \texttt{flux\_density\_wavelength} & erg/cm$^2$/s/\AA & \texttt{Sed.flam} ($F_\lambda$) \\
    \texttt{mass\_rate} & Msun/yr & \texttt{BlackHoles.accretion\_rates} ($\dot{M_\bullet}$) \\
    \hline
  \end{tabular}
  \tablecomments{\texttt{Category} is the overarching label used within Synthesizer for a set of variables sharing a common unit. The second column shows the unit for each category. The final column provides example attributes from Synthesizer with the equivalent mathematical notation.}
\end{table*}

All non-dimensionless attributes on \synthesizer objects are in reality \texttt{Quantity} objects. The \texttt{Quantity} is a descriptor class that defines how attributes with units are `get' and `set', automating the application of the unit system. The user will never instantiate a \texttt{Quantity} themselves.
\texttt{Quantity} objects carry a reference to the \texttt{Units} singleton, giving them access to the global unit system. When a \texttt{Quantity} is set on an object, its value (without units) is stored on the object after conversion to the appropriate unit in the unit system when necessary. When the attribute is retrieved from an object (i.e. \texttt{foo.bar}) the previously set value is returned, including units. If instead the private version of the attribute is requested (i.e. \texttt{foo.\_bar}) then the \textit{unitless} version is returned. Internally, calculations are predominantly done using the private unitless form to avoid unit overhead unless unit safety is required (such as when working with user input).

We also include a modified version of the \texttt{accepts} decorator implemented in \unyt. This decorator defines the expected units for arguments into the decorated function, and raises an error if the units are either missing or not compatible. Our modified version builds on this behaviour by folding in our unit system and converting all arguments with units to the expected unit, after checking that units are provided and compatible. This guarantees that the user knows what has been passed into a function at the point of use, and captures unit errors early.

\section{Abundances}
\label{sec:abundances}

\textsc{Synthesizer} offers users the flexibility and freedom to set the abundances of different elements (H, He and metals) and dust. This feature is particularly useful for interfacing with photoionisation codes (such as \textsc{cloudy}, \textsc{Mappings},) which requires the elemental abundances to be set with respect to hydrogen.
Within \synth, this can be set by using reference abundance pattern available in the literature. Currently, three abundance patterns are supported, based on \cite{Asplund2009}, \cite{GalacticConcordance2017}  and \cite{gutkin_modelling_2016}. Additional flexibility is provided by allowing users to scale the elemental abundances directly (e.g. for C, N, O), based on their ratios (such C/O or N/O) or alpha-enhancement ([$\alpha$/Fe]). These scalings can also follow observed empirical relations from the literature, such as those in \cite{dopita2006_scaling} or \cite{GalacticConcordance2017}. 

\synth\ also provides the option to set the depletion of refractory elements into dust. The main purpose is to interface with photoionisation models, similar to the elemental abundances. By applying a chosen depletion pattern, the abundance of each refractory element is reduced accordingly. Currently supported depletion patterns include those presented in \cite{Jenkins2009_depletion} \cite[implemented for \textsc{cloudy} v23.01 in][]{Gunasekhara2022_depletion}, the fiducial \textsc{cloudy} depletion (see Table 7.8 in Hazy1) and the one used in \cite{gutkin_modelling_2016}. Based on the amount of elemental depletion onto dust, \synth\ can interface with \textsc{cloudy} and set the mass in different dust grains accordingly.

These helper functions are designed to provide a framework for setting elemental abundances motivated by previous literature. 
While they are not the only possible scalings, it is important to remain consistent once a particular scaling is chosen. Inconsistencies can lead to abundance fractions that do not sum to unity.

\section{Creating Your Own Grids}
\label{sec:grids_create}
Advanced users can create their own \textsc{Synthesizer} grids.
These can be intrinsic grids of stellar emission, generated from stellar population synthesis models, or grids post-processed through photoionisation codes such as cloudy.
The code for creating custom grids is contained in an independent repository, \url{https://github.com/synthesizer-project/grid-generation}.
You will need a working installation of \synth\ for these scripts to work, as well as other dependencies for specific codes (e.g. \textsc{Cloudy}, python-FSPS).
Grids should follow the naming convention where possible, see \sec{sps_grids}.
Please see \app{abundances} for details on how to modify the chemical abundance pattern of gas, stars and dust using the `abundances' object, and use this when running cloudy.

\section{Simulation Front Ends}
\label{sec:front_ends}

\textsc{Synthesizer} contains front ends for a number of different simulation codes.
These include the cosmological hydrodynamic simulations \textsc{Eagle} \citep{schaye_eagle_2015,crain_eagle_2015}, \textsc{Simba} \citep{dave_simba:_2019}, \textsc{Illustris-TNG} \citep{pillepich_first_2018,nelson_first_2018,springel_first_2018,naiman_first_2018,marinacci_first_2018}, \textsc{BlueTides} \citep{feng_formation_2015,feng_bluetides_2016}, \textsc{Flares} \citep{lovell_first_2021,vijayan_first_2021} and CAMELS \citep{villaescusa-navarro_camels_2021}.

For simulations that are not listed above, the methods developed for these simulations can be used as a starting point to develop custom data loaders, leveraging the \texttt{galaxy} object and individual component initialisation routines and \texttt{unyt} definitions.
We encourage users to develop and contribute new loaders for their simulation of choice to the code base.

\section{Naming conventions}
\label{sec:em_naming}

\synth\ enables the generation of many different spectra which are associated with Galaxy objects or their components through \texttt{EmissionModel}'s.
These spectra are given standardised labels that reflect their origin and the masks that have been applied (though custom labels can be provided).
These labels are in most cases derived from the Cloudy naming conventions, for consistency. 
\fig{flowchart} shows how these different spectra are typically generated and related by an \texttt{EmissionModel}.

Our standard naming system, which is used in the premade \texttt{EmissionModels}, is described below:
\begin{itemize}
    \item \texttt{incident} spectra are the `raw' spectra produced by a component, such as a stellar population or AGN. In the context of stars, incident spectra are the spectra that are produced by a stellar population synthesis code, and equivalent to the “pure stellar” spectra. The motivation for the name is that these are the spectra that serve as the source for photoionisation modelling, which we will explain below. 

    \item \texttt{transmitted} spectra are the incident spectra that are transmitted through the gas in the photoionisation modelling. Functionally, the main difference between transmitted and incident is that the transmitted has little flux below the Lyman-limit, since this has been absorbed by the gas. This depends on the escape fraction (\texttt{fesc}).

    \item \texttt{nebular} is the combination of continuum and line emission predicted by the photoionisation model. This depends on the escape fraction (\texttt{fesc}).

    \item \texttt{reprocessed} is the emission which has been reprocessed by the gas. This is the sum of \texttt{nebular} and \texttt{transmitted} emission.

    \item \texttt{escaped} is the incident emission that escapes reprocessing by gas. This is \texttt{fesc} $\times$ \texttt{incident}. 
    This is not subsequently affected by ISM dust.

    \item \texttt{intrinsic} is the sum of the \texttt{escaped} and \texttt{reprocessed} emission, essentially the emission before ISM dust attenuation.

    \item \texttt{attenuated} is the \texttt{reprocessed} emission with attenuation by dust in the ISM.

    \item \texttt{emergent} is the combined emission including dust attenuation and is the sum of \texttt{attenuated} and \texttt{escaped}. Note that this does not include thermal dust emission, so is only valid from the UV to near-IR.
     
    \item \texttt{dust\_emission} is the thermal dust emission, typically calculated using an energy balance approach, and assuming a dust emission model.

    \item \texttt{total} is the sum of \texttt{attenuated} and \texttt{dust\_emission}, i.e. it includes both the effect of dust attenuation and dust emission.

    \item For two component dust models \cite[e.g.][or \texttt{BimodalPacmanEmission}]{charlot_simple_2000}
    we also generate the individual spectra of the \texttt{young} and \texttt{old} component. This includes \texttt{young\_incident}, \texttt{young\_nebular}, \texttt{young\_attenuated}, etc.; \texttt{young} and \texttt{old} are equivalent to \texttt{total} for the young and old components.
\end{itemize}

All pre-made models follow these naming conventions, and we encourage users to employ the same conventions where possible.






\section{Fiducial Emission Models}
\label{sec:fiducial_emodel}

Below we describe in detail some of the default emission models supplied with \textsc{Synthesizer}.
These can be modified or combined for any particular use case, or new emission models can be built from scratch.
More details on the modelling of dust in particular is provided in \sec{dust}.



\subsection{Screen Attenuation Model}

A screen attenuation model is primarily a dust attenuation emission model, that assumes a uniform screen of dust in front of the stellar component that attenuates all emission equally, using the same attenuation curve.
This model is analogous to a number of models in the literature, such as those applied to semi-analytic models \citep[e.g.][]{lacey_tidally_1993} and hydrodynamic simulations \citep[e.e.][]{trayford_colours_2015}.

\subsection{Charlot \& Fall 2000 Attenuation Model}

The \cite{charlot_simple_2000} attenuation model (CF00 hereafter) is an extension to the screen attenuation model that includes two separate screens.
The first attenuates all stars, and mimics the impact of dust in the diffuse interstellar medium.
The second screen attenuates only young stars, and is analogous to dust in star forming HII regions, which provides additional attenuation to young stellar populations.

The CF00 model has been applied to both semi-analytic models \citep[e.g.][]{somerville_galaxy_2012} and hydrodynamic simulations \citep[e.g.][]{nagamine_massive_2005,trayford_colours_2015,vijayan_first_2021,lovell_learning_2024}.

\subsection{The Pacman Model}

\begin{figure*}
\centering
	\includegraphics[width=0.8\textwidth]{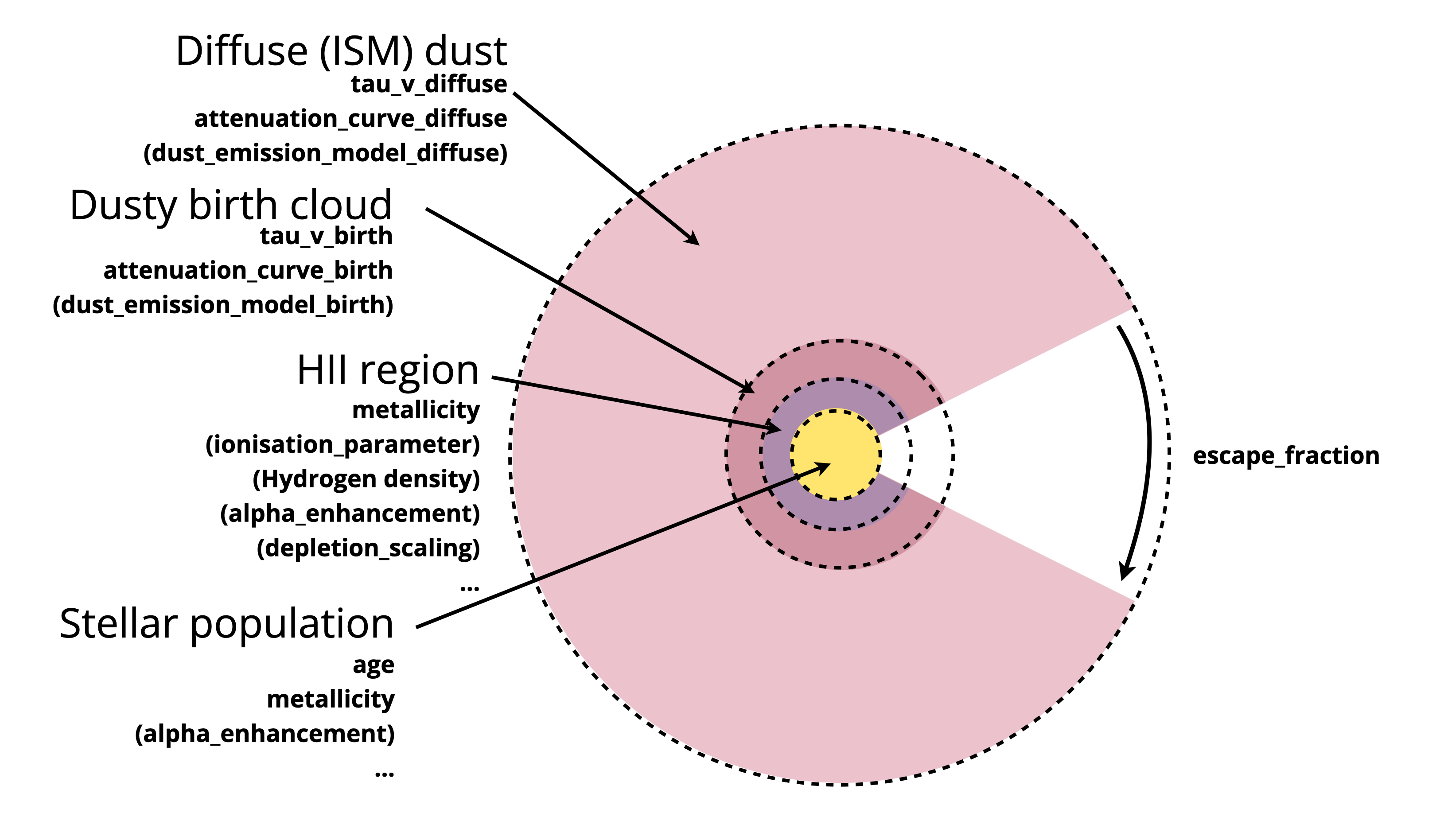}
    \caption{Diagram showing the various components of the Pacman emission model. This includes the contribution of a source stellar population, HII region photoionisation, birth cloud dust attenuation and diffuse ISM dust attenuation, as well as the escape of ionising radiation given an assumed escape fraction.}
    \label{fig:pacman}
\end{figure*}

The Pacman Model is a generalised model that fully encapsulates the complexity of emission from young star forming regions.
Star forming regions have highly complex geometries of stars, gas and dust, which leads to leakage of radiation from ionising sources.
The `pacman' simplifies this to an \textit{escaped} fraction, analogous to a single avenue for radiation to escape out of the cloud, which can be visualised as a slice out of a circle of gas (the `mouth' of the pacman).

The intrinsic emission from the Pacman model can be attenuated using the screen models described above, or using a line-of-sight attenuation model (see \sec{los}).
\fig{pacman} shows a diagram outlining the Pacman model.


\subsubsection{UnifiedAGN}
\label{sec:em_agn}

As noted in \S\ref{sec:grids_agn} \synth\ also adopts a flexible approach to modelling the emission from AGN.
The simplest model simply scales a template by the bolometric luminosity (calculated from the accretion rate and radiative efficiency).
A simple extension here is the use of a grid of templates, for example where the SED depends on SMBH mass or accretion rate.

We also implement a more sophisticated model, UnifiedAGN, which self-consistently combines the emission from the disc, broad and narrow line regions, and a dusty torus. The point of this model is to replicate observations of classical Type I and II AGN.
This model, and the photoionisation modelling that supports it, is comprehensively described in Wilkins et al. \emph{in-prep} and Vijayan et al. \emph{in-prep}.

First, the disc emission is described by the choice of disc emission model.
At present several options are available as described in \S\ref{sec:grids_agn}.
The disc model is assigned based on the physical properties of the AGN, either the bolometric luminosity or the mass, accretion rate, inclination, and \emph{potentially} spin.
Unless there is an \emph{ab initio} reason to do so the inclination is set to a random value. 

The broad and narrow line regions are assumed to be illuminated by the entire disc and to emit isotropically.
Both regions are assumed to only reprocess some fraction ($f_{\rm cov, NLR}$, $f_{\rm cov, BLR}$) of the disc, with the remainder ($f_{\rm cov, NLR} + f_{\rm cov, BLR}$) escaping. 

A key element of this model is a thick dusty torus, which is assumed to be aligned with the disc. When the inclination of the disc is smaller than the opening angle of the torus $\theta$ the direct emission from both the disc and BLR is blocked, leaving only the emission from the NLR and torus visible, i.e. comparable to a classical Type II AGN.
When the inclination is larger the disc and BLR are both also visible resulting in a Type I AGN. The torus is assumed to reprocess some fraction (determined by its opening angle) of the emission from the AGN. This is re-emitted assuming a choice of dust-emission template. 

Finally, the emission from the individual components - or the total emission - can also be attenuated by intervening dust in the interstellar medium.

\section{Generating spectra for Semi-analytic models}
\label{sec:sam_spectra}

\begin{figure*}
	\includegraphics[width=\textwidth]{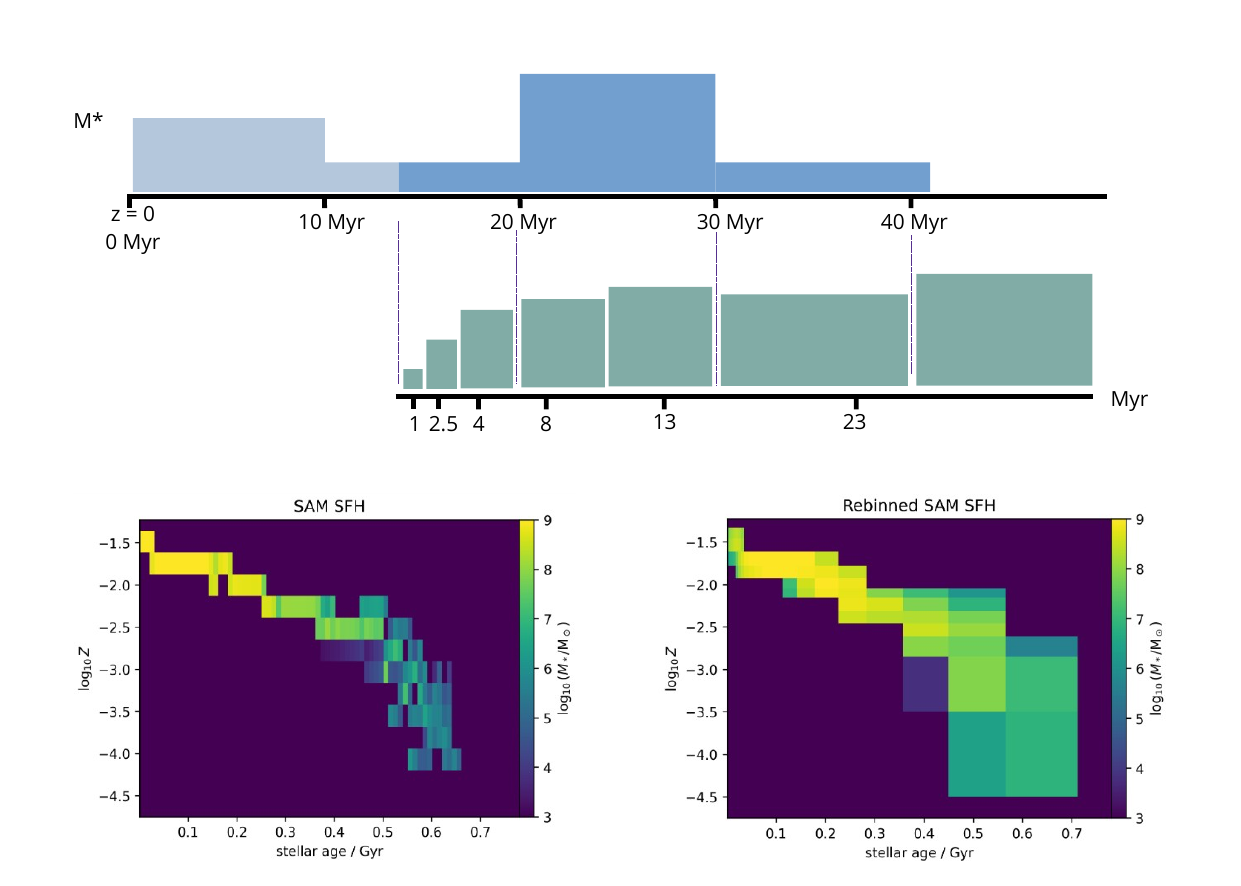}
    \caption{Top: Illustration of the grid assignment scheme for a SAM output on a regular grid. The top line shows a galaxy star formation history distributed on this regular grid. The bottom line shows the SPS grid outputs, and the mass assigned to each point on the grid.
    Bottom left: example distribution of stellar mass in an SC-SAM output grid, after converting the age axis from lookback time to stellar age. Bottom right: distribution of stellar mass in the same SC-SAM galaxy after rebinning the output grid to the SPS grid.}
    \label{fig:sam_sfh}
\end{figure*}

Semi-analytic models (SAMs) are a special case when generating spectra, depending on how their star formation histories are output to disk.
Below we describe a simple output format, whereby a fixed grid of output times are defined at runtime, in terms of the lookback time from $z = 0$, and the SFH is output on this fixed grid when the SAM is run.
The mass formed at each point is saved, along with the averaged metallicity.
The SFH of a galaxy at any chosen redshift is then given by truncating this grid up to the output time.
This is the output format used in the Santa-Cruz SAM \citep[SC-SAM;][]{somerville_star_2015}, and we provide a method for reading SFHs from this simulation with \textsc{Synthesizer}.
However, we stress that the same method can be applied to other SAMs with similar output formats (see \app{front_ends} for details on other simulation front ends supported to date).

Before creating a galaxy object from an output grid, it is necessary to translate the axes of the output grid into that of the desired SPS grid.
The first step of this process is to define the age axis of the output grid in terms of stellar age rather than lookback time.
The resulting output grid is then rebinned to the SPS grid by considering the overlap between cells in the output grid and cells in the SPS grid.
In essence, the stellar mass in each output grid cell is assigned to overlapping SPS grid cells in proportion to the area of overlap. We ensure that stellar mass is conserved for each output grid cell.
This method requires grid cells to have an associated area, and hence defined boundaries.
In the example shown in Figure \ref{fig:sam_sfh}, we have chosen to define metallicity bin edges as being equidistant from bin centres in log-space, while the stellar age bin edges are equidistant from bin centres in Gyr.

The age bins of the SPS grid are much more finely spaced at young stellar ages compared to the SC-SAM output grid.
Thus, at stellar ages of $<0.05\ {\rm Gyr}$, there is less stellar mass per bin in the rebinned SFH (bottom panel) compared to the original SFH (top panel).
These young age bins also require special treatment at $z>0$, where the output time may lie between two bins.
In this case, if the first bin directly \textit{before} the output time has a non-zero SFR, the bin is truncated and assigned a weight proportional to the SFR in the previous bin, and the fraction of the bin covered by this output time.

\section{Parametric Stellar morphologies}
\label{sec:parametric_stellar_morph}

\begin{figure}
	\includegraphics[width=\columnwidth]{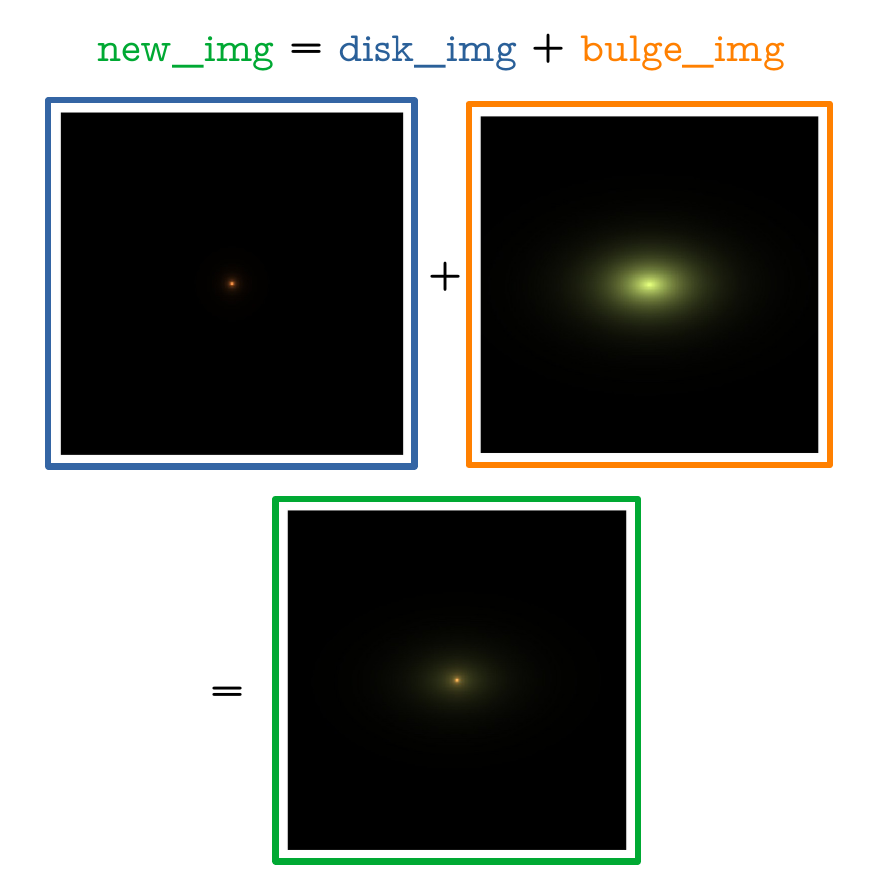}
    \caption{Example of defining a parametric disk morphology (top right), described using a S\'{e}rsic 2D profile, and a bulge morphology (top left), described using a Gaussian 2D profile.
    These are then coupled to a SFZH and a grid, and used to generate RGB images from the UVJ bands (ordered by wavelength appropriately).
    The bottom panel shows the combination of these images.}
    \label{fig:parametric_morph}
\end{figure}

Parametric stars objects can optionally define a morphology, describing the spatial distribution of the stars through parametric forms.
Currently supported morphologies include a 2D Gaussian, a 2D S\'{e}rsic profile, and a point source.
These morphologies can be coupled directly with individual stellar component star formation and metal enrichment histories to predict their spectra, and all derived observables, such as photometry.
Stellar morphology objects can also be easily combined by using the overloaded addition operator.

A number of different parametric morphologies are provided, with arguments that depend on whether the morphology is defined in physical cartesian coordinates, or in angular coordinates in the sky plane.
These arguments must be defined with unyt units.
The \texttt{Morphology} class can convert between cartesian and angular coordinates, but only if a cosmology object and redshift of the galaxy is provided.

Different parametric morphology components can be added together, to mimic the contribution of different stellar populations that are distinct spatially; for example, a bulge and disk can be combined to mimic a classical spiral galaxy.
This can be achieved using the overridden addition operator on two image collections.
An example of this is shown in \fig{parametric_morph}, showing a disk and bulge morphology, after coupling to a star formation and metal enrichment history (SFZH) and grid, and how these can be used to generate RGB images from multi-band photometry.

\section{Property Maps}
\label{sec:property_maps}
\textsc{Synthesizer} provides helper methods for making commonly used maps of physical properties.
Unlike the methods used to make collections of photometric images, these map making methods return individual \texttt{Image} objects.
As for Images, property maps can be smoothed over the particle kernel, or return a simple summed histogram.
Note that some properties are summative (e.g. mass) while others are mass weighted averages (e.g. metallicity and age).
In the latter case, maps of the property multiplied by the mass are then normalised by the mass map,
\begin{equation}
    F(x,y) = \frac{\sum_j m_j W(r_j , h_j) F_j }{\sum  m_j W(r_j , h_j)} \;\;.
\end{equation}
where $F$ is the property of interest for particle $j$, $m_j$ is the normalisation property, and $W$ is the particle kernel.
Property maps provided by default include stellar mass, metallicity (total or mass-weighted), age (mass or luminosity weighted), star formation rate and specific star formation rate (either instantaneous or given some timescale).
Users can also define their own property maps using any property or combination of properties stored on the particle data.
These can optionally be normalised by a secondary property (e.g. the mass).

A number of example smoothed property maps are shown in \fig{mosaic}, including an example of a custom map constructed using the H-$\alpha$ emission of each star particle.

\section{Dust Modelling}
\label{sec:dust}

Whilst the impact of dust in an \emodel is treated simply as a \textit{transformation}, dust has a number of unique aspects that are worth describing in detail.
These include a number of fiducial attenuation curves and dust emission models provided within synthesizer, and the line-of-sight (LoS) attenuation model.

\subsection{Dust Attenuation Curves}
\label{sec:dust_curves}

\begin{figure}
	\includegraphics[width=\columnwidth]{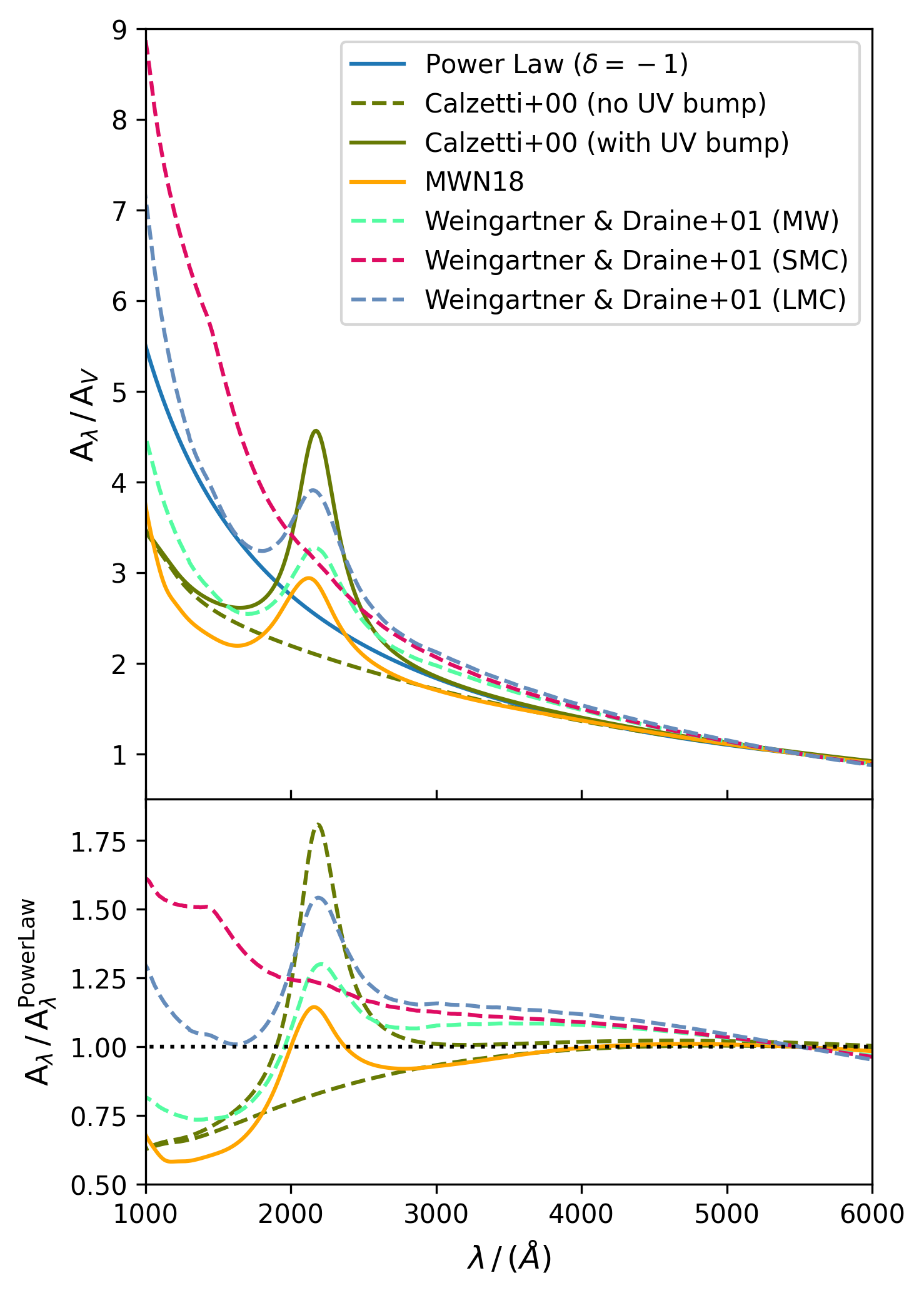}
    \caption{Attenuation curves available in \textsc{synthesizer}. 
	We show a simple power law, with slope $\delta = -1$, a \protect\cite{calzetti_dust_2000} model with and without the 2175$\AA$ bump feature, a Milky Way (MW) curve from \protect\cite{narayanan_theory_2018}, and the MW, Large Magellanic Cloud (LMC) and Small Magellanic Cloud (SMC) curves from \protect\cite{weingartner_dust_2001}.
	Not shown is the \protect\cite{li_dust_2008} parametric model, which can also provide predictions for the Calzetti$+$00, MW, LMC and SMC curves, as well as the flexibility to model other attenuation curves through its 4 parameter form.
	The bottom panel shows the ratio of the attenuation in each model with respect to the power law model.
    }
    \label{fig:attenuation_curves}
\end{figure}

\textsc{Synthesizer} includes a number of attenuation / extinction curves\footnote{We will use the term \textit{attenuation curves} throughout; for the difference see Figure 1 in \cite{salim_dust_2020}} (shown in \fig{attenuation_curves}):
\begin{itemize}
	\item \textbf{PowerLaw}: A simple power-law attenuation curve.
	\item \textbf{Calzetti2000}: The \cite{calzetti_dust_2000} attenuation curve (with an optional UV bump from \citealt{noll_analysis_2009}).
	\item \textbf{MWN18}: A Milky Way attenuation curve, defined in \cite{narayanan_theory_2018}.
	\item \textbf{GrainsWD01}: A dust grain attenuation curve from \cite{weingartner_dust_2001} with models for the Milky Way, LMC, and SMC (and more defined in \citealt{weingartner_dust_2001}).
	\item \textbf{ParametricLi08}: A parametric and empirically derived attenuation curve implemented in \cite{li_dust_2008}, including parameters from \cite{markov_dust_2023} and \cite{markov_evolution_2025} extending it to $z = 12$.
\end{itemize}

Attenuation curves can be instantiated directly, or attached to an \texttt{EmissionModel} (see \sec{emission_models}).
Each can return the transmission or attenuation as a function of wavelength, given a V-band optical depth, $\tau_{\rm V}$.

\subsection{Line of Sight Attenuation Model}
\label{sec:los}

\begin{figure}
	\includegraphics[width=\columnwidth]{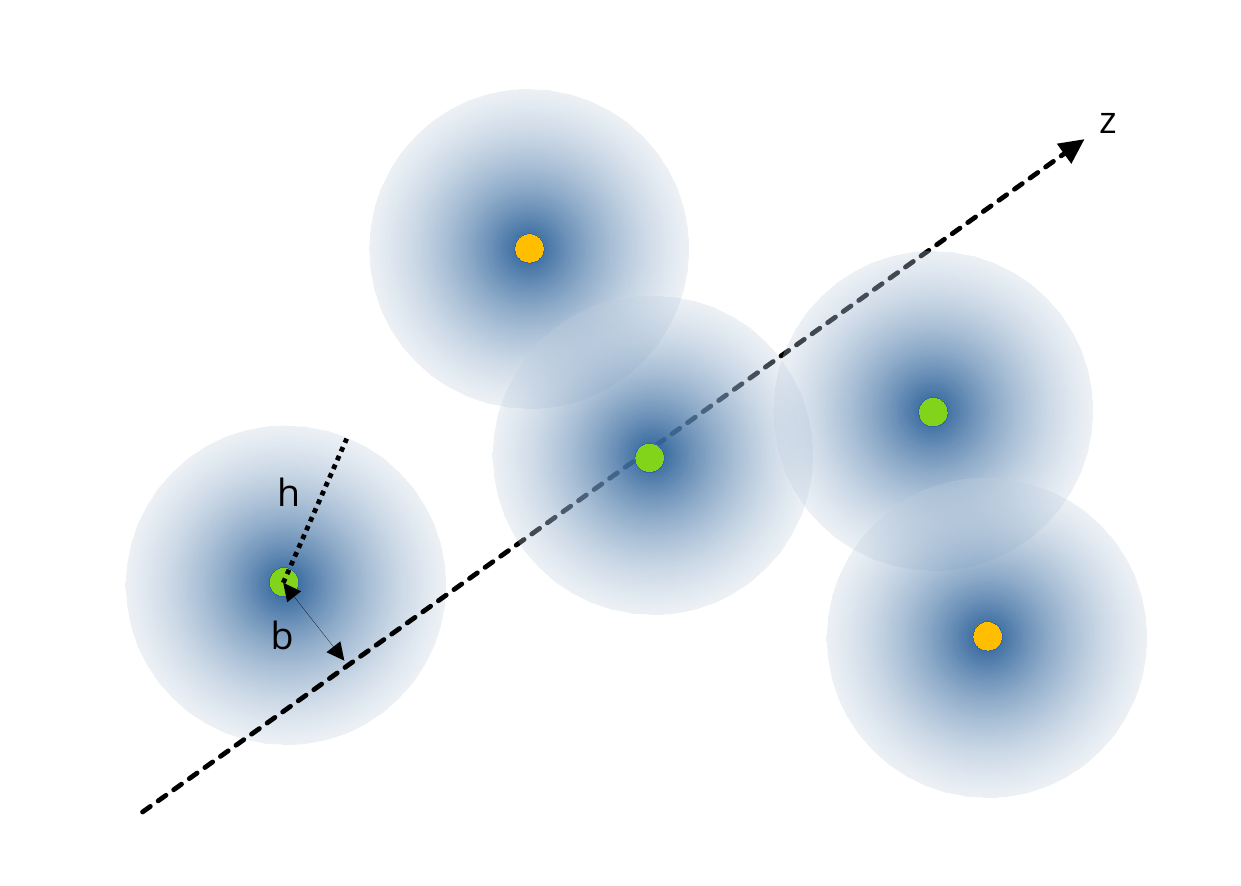}
    \caption{Diagram demonstrating the Line of Sight (LoS) attenuation model.
	Particles that intersect the chosen line of sight direction ($z$) are labelled with green points.
	The blue represents the kernel of each gas/dust particle, with the smoothing length given by $h$.
	The impact parameter $b$ of each particle that intersects the line of sight is used to obtain the column density through a pre-computed lookup table.
	Particles that do not lie along the chosen LoS ($b > h$) are coloured orange.}
    \label{fig:los}
\end{figure}


The Line of Sight (LoS) attenuation model is a more sophisticated model for the diffuse interstellar dust attenuation that can be computed for a Particle Galaxy object, by taking into account the full three dimensional star-dust or black hole-dust geometry.
Where the mass and distribution of dust is self-consistently predicted by the simulation \citep[e.g.][]{aoyama_galaxy_2017,mckinnon_simulating_2017,gjergo_dust_2018,li_dust--gas_2019} these values can be used directly to calculate the dust column density towards each star or black hole particle, assuming a given viewing direction.
Where this information is not available, the distribution of gas-phase metals can be used as a proxy for the dust distribution \citep[e.g.][]{wilkins_properties_2017,vijayan_first_2021}.
In this approach the column density of gas-phase metals along the line-of-sight to each emitting particle is computed by summing the contributions from each gas particle, assuming a 3D kernel function.
This column density is then related to the dust column by assuming some Dust-to-Metal (DtM) ratio.
This DtM can be fixed, or depend on integrated properties of the galaxy, such as the gas phase metallicity \citep{remy-ruyer_2014} or redshift \citep{vogelsberger_high-redshift_2020}.
In both cases this approach requires details on the star particle distribution as well as the gas / dust.
Subgrid dust e.g. within star forming clouds, can be accounted for within each star particle source.

In detail, we treat each star particle as a point source. The viewing angle can be arbitrarily chosen, but in the below we assume this is along the $z$-axis.
The metal column density, $\Sigma(x, y)$, integrated along the LoS, is related to the dust optical depth as,
\begin{align}
    \tau_{\rm ISM,V} (x, y) = {\rm DtM}\, \kappa_{\rm ISM} \, \Sigma(x, y) \;,    
\end{align}


where $\tau_{\rm ISM,V}$ is the optical depth in the V-band (5500 \AA), DtM is the dust-to-metal ratio, and $\kappa_{\rm ISM}$ is a normalisation parameter that encodes the properties of the dust, such as the composition, grain size and shape.
This latter parameter is often calibrated, e.g. to the UV luminosity function in high redshift scenarios \citep{vijayan_first_2021}.
$\Sigma(x, y)$ is the metal column density; where the dust properties are self consistently predicted, this simplifies to $\tau_{\rm ISM,V} = \Sigma_{\rm dust}(x, y)$, using the direct dust column densities.
The DtM value of a given galaxy can be fixed (typically $\rm DtM=0.3$), linked to the integrated physical parameters of the galaxy \citep[e.g.][]{vijayan_detailed_2019}, or allowed to vary particle-by-particle, allowing for spatially varying dust properties.

$\Sigma$(x, y) is obtained by integrating the density field of particles that intersect the line of sight, assuming a smoothing kernel for each particle in the simulation.
This is controlled for each particle by the impact parameter, $b$.
The LoS metal column density ($\Sigma\,(x,y)$) is calculated as follows:
\begin{equation}\label{eq: Zlos}
\Sigma\,(x,y) = 2 \sum_i Z_i m_i \int_{0}^{\sqrt{h_i^2-b_i^2}}W(r, h_i)dz\,;\, r^2 = b_i^2 + z^2\:,
\end{equation}
where $i$ denotes each gas particles along the LoS, $Z$ and $m$ describe the metallicity and mass of each particle, and $W(r, h_i)$ describes the smoothing kernel for each particle, with smoothing length $h$.
The impact parameter is typically normalised by the smoothing length, allowing the generation of pre-computed tables of the LoS column density.
These can then be used to compute the dust column density for arbitrary values of smoothing length and impact parameter for all particles in a given Particle Galaxy object.

\fig{los} shows a simplified example of the LoS approach for a single star particle, and a collection of `dust' particles.

The optical depth in the V-band is linked to other wavelengths as
\begin{equation}\label{tau_lambda}
	\tau_{\lambda} = \tau_{\textrm{ISM}} \times\, k_{\lambda},
\end{equation}
where k$_{\lambda}$ can be any of the V-band normalised dust attenuation curve described in Section \ref{sec:dust_curves}.

\begin{figure}
	\includegraphics[width=\columnwidth]{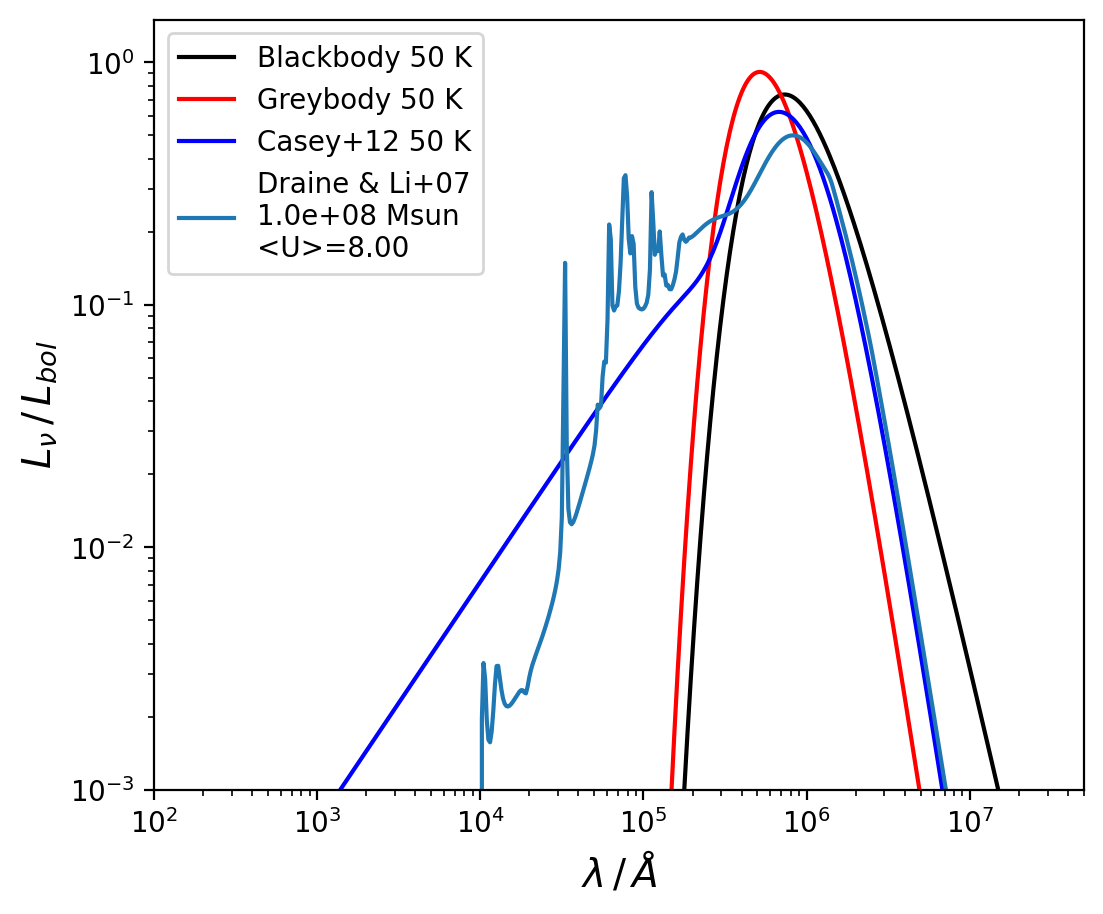}
    \caption{Dust emission models. A Blackbody, modified Greybody and the \protect\cite{casey_far-infrared_2012} model, all assuming a dust temperature of 50 K, as well as the \protect\cite{draine_infrared_2007} model for a dust mass of $10^8 \; {\rm M_{\odot}}$.}
    \label{fig:dust_emission}
\end{figure}

\subsection{Dust Emission}
\label{sec:dust_emission}

\textsc{Synthesizer} provides a number of parametric models for the thermal dust emission, including a pure blackbody, greybody (non-unity emissivity), the \cite{casey_far-infrared_2012} model (separate parameterisation for the mid-IR and far-IR) and the \cite{draine_infrared_2007}  model\footnote{These pre-computed grids are made available through the synthesizer command line interface, \texttt{synthesizer-download --dust-grid}}
\cite[updated with the 2014 grids, and using the parametrised dust luminosity per unit dust mass using the relation from][]{Magdis2012}.
By default, the bolometric luminosity of the spectrum produced by the dust emission model is normalised to 1. This is then scaled by the dust bolometric luminosity computed assuming energy balance by taking the difference between the intrinsic and dust attenuated bolometric luminosity. However, this is not the case for the \cite{draine_infrared_2007} model, where there is a fixed relationship between the dust luminosity and the dust mass. 
The impact of CMB heating can be accounted for by providing the redshift, which is used to estimate the temperature, $T_{\rm CMB} = 2.73 \times (1 + z)$.
Some example dust emission spectra from different models are shown in \fig{dust_emission}.

\section{IGM Absorption}
\label{sec:igm_absorption}

\begin{figure}
    \includegraphics[width=\columnwidth]{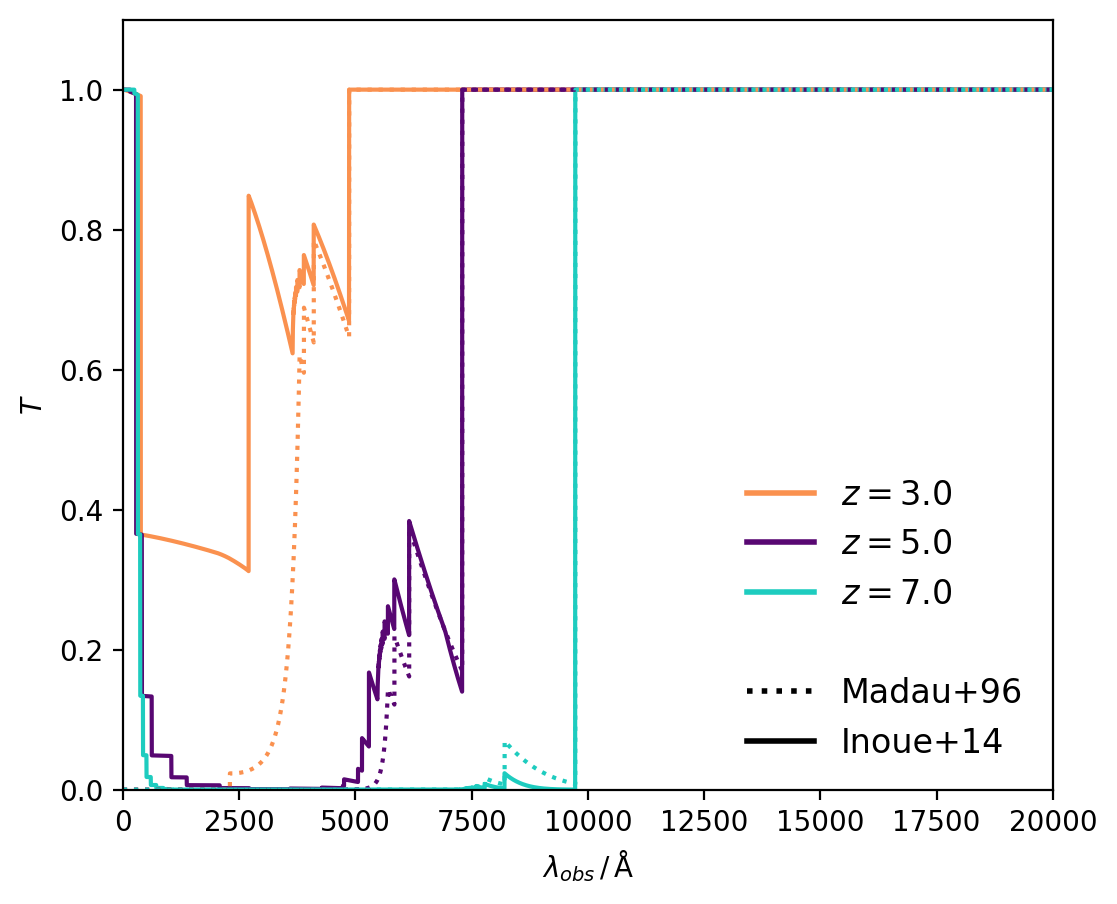}
    \caption{IGM absorption produced by the \protect\cite{madau_21_1997} and \protect\cite{inoue_updated_2014} models at $z = 3$.}
    \label{fig:igm_transmission}
\end{figure}

Neutral hydrogen in the intergalactic medium (IGM) attenuates the light from distant galaxies, even after reionisation.
\textsc{Synthesizer} provides two analytic forms for this IGM absorption, \cite{madau_21_1997}, which assumes a power-law relationship between the absorption and the redshift, and \cite{inoue_updated_2014}, which includes the effects of the Lyman-$\alpha$ forest and Lyman-limit systems.
\fig{igm_transmission} shows both of these forms from $z = 3$ to $7$.

\section{Smoothing Kernels}
\label{sec:kernels}

We encode a number of common smoothing kernels for use in various parts of the package, including the LoS optical depth calculation (see \sec{dust}) and imaging.
These include the \textsc{Anarchy} SPH scheme, used in \textsc{Eagle} and \textsc{Flares}  \citep[see][for more details]{schaller_eagle_2015}.
Users can define their own custom kernels as required.


\section{Configuration Options}

\textsc{Synthesizer} uses C++ extensions for many of the computationally heavy background operations.
When installing it’s possible to enable special behaviours and control the compilation of these C++ extensions by defining certain environment variables:
\begin{itemize}
    \item \verb|WITH_OPENMP| controls whether the C++ extensions will be compiled with OpenMP threading enabled. This can be set to an arbitrary value to compile with OpenMP or can be the file path to the OpenMP installation directory (specifically, the directory containing the include and lib directories). This later option is useful for systems where OpenMP is not automatically detected, such as OSX when libomp has been installed with homebrew.
    \item \verb|ENABLE_DEBUGGING_CHECKS| turns on debugging checks within the C++ extensions. These (expensive) checks are extra consistency checks to ensure the code is behaving as expected. Turning these on will slow down the code significantly.
    \item \verb|RUTHLESS| turns on almost all compiler warnings and converts them into errors to ensure the C++ code in \textsc{Synthesizer} is clean. Note that this does not include the \texttt{--pedantic} flag, because Python and Numpy themselves do not adhere to these rules, and thus do not compile using \texttt{gcc} and this flag.
    \item \verb|ATOMIC_TIMINGS| turns on low level timings for expensive computations. This is required to use time related profiling scripts.
    \item \verb|CFLAGS| allows the user to override the flags passed to the C++ compiler. Simply pass your desired flags to this variable.
    \item \verb|LDFLAGS| allows the user to override the directories used during linking.
    \item \verb|EXTRA_INCLUDES| allows the user to provide any extra include directories that fail to be automatically picked up.
\end{itemize}

\end{appendix} 

\end{document}